\documentclass[11pt]{article}
\usepackage{epsfig}
\font\cmr=cmr7

\newcommand{\be}{\begin{equation}}
\newcommand{\ee}{\end{equation}}
\newcommand{\bea}{\begin{eqnarray}}
\newcommand{\ena}{\end{eqnarray}}
\newcommand{\vs}[1]{\rule[- #1 mm]{0mm}{#1 mm}}

\newcommand{\rs}{{\sqrt s}}
\newcommand{\pT}{{p_{_T} }}
\newcommand{\dsig}{  {d \sigma 
		\over {d {\vec p}_{_T} d \eta}}}
\newcommand{\dsigd}{  {d \sigma^{\hbox{\cmr dir}} 
		\over {d {\vec p}_{_T} d \eta}}}
\newcommand{\dsigb}{  {d \sigma^{\hbox{\cmr brem}} 
		\over {d {\vec p}_{_T} d \eta}}}
\newcommand{\dsigij}{  {d {\widehat \sigma}_{ij} 
		\over {d {\vec p}_{_T} d \eta}}}
\newcommand{\dsigijk}{  {d {\widehat \sigma}_{ij}^k 
		\over {d {\vec p}_{_T} d \eta}}}
\newcommand{\kd}{  K^{\hbox{\cmr dir}} }
\newcommand{\kb}{  K^{\hbox{\cmr brem}} }
\newcommand{\MSB}{{\overline {MS}}}
\newcommand{\LMS}{\Lambda_{_{_\MSB}}}
\newcommand{\as}{\alpha_{s}(\mu)}
\newcommand{\alfspi}{ {\alpha_s(\mu) \over 2 \pi} }

\newcommand{\asmfpi}{ {\alpha_s(M_{_F}) \over 2 \pi} }
\newcommand{\NP}[1]{Nucl.\ Phys.\ {\bf #1}}
\newcommand{\PL}[1]{Phys.\ Lett.\ {\bf #1}}

\newcommand{\PR}[1]{Phys.\ Rev.\ {\bf #1}}
\newcommand{\PRL}[1]{Phys.\ Rev.\ Lett.\ {\bf #1}}
\newcommand{\ZPH}[1]{Z.\ Phys.\  {\bf #1}}

\newcommand{\EPJ}[1]{Eur.\ Phys.\ J.\ {\bf #1}}

\begin{document}

\renewcommand{\thefootnote}{\fnsymbol{footnote}}
\newpage
\setcounter{page}{0}

\vs{10}

\begin{center}
{\Large {\bf{A CRITICAL PHENOMENOLOGICAL STUDY OF}}} \\
\vspace{0.5 cm}
{\Large {\bf{INCLUSIVE PHOTON PRODUCTION}}} \\
\vspace{0.5 cm}
{\Large {\bf{IN HADRONIC COLLISIONS}}} \\
\vspace{0.7 cm}
{\large P.~Aurenche$^1$, M.~Fontannaz$^2$, J.Ph.~Guillet$^1$, 
B. Kniehl$^3$, E.~Pilon$^1$, M.~Werlen$^1$} \\
\vspace{0.7 cm}
\end{center}
\begin{itemize}
\item[1.] {\em Laboratoire de Physique Th\'eorique LAPTH,
\footnote{URA 14-36 du
CNRS, associ\'ee \`a l'Universit\'e de Savoie.}\\
B.P. 110, F-74941 Annecy-le-Vieux Cedex, France} \\
\item[2.] {\em Laboratoire de Physique Th\'eorique et Hautes Energies LPTHE,
\footnote{UMR 63 du CNRS.}\\
B\^at. 211, Universit\'e de Paris-Sud, F-91405 Orsay Cedex, France} \\
\item[2.] {\em Max-Planck-Institut f\"ur Physik 
(Werner-Heisenberg-Institut), \\
F\"ohringer Ring 6, D-80805 Munich, Germany} \\
\end{itemize}

\vs{10}

\centerline{ \bf{Abstract}}
\vs{3}

We discuss fixed target and ISR inclusive photon production and attempt
a comparison between theory and experiments.  The dependence of the
theoretical predictions on the structure functions, and on the
renormalization and factorization scales is investigated. The main
result of this study is that the data cannot be simultaneously fitted
with a single set of scales and structure functions. On the other hand,
there is no need for an additional intrinsic $k_{_T}$ to force the
agreement between QCD predictions and experiments, with the possible
exception of one data set. Since the data cover almost overlapping
kinematical ranges this raises the question of consistency among data
sets. A comparative discussion of some possible sources of experimental
uncertainties is sketched.

\vfill
\rightline{hep-ph/9811382}
\rightline{LAPTH-704/98}
\rightline{LPTHE-Orsay/98/69}
\rightline{MPI/PhT/98-083}
\rightline{November 1998}


\newpage
         
\indent

Despite many years of intense 
experimental~\cite{EXP,R806,WA70,WA70pi,R110,R807,E70693,E706,UA6,VW} and 
theoretical~\cite{FRIT,ABDFS,ACFGP,GV} efforts the inclusive
production of prompt photons in hadronic collisions does not
appear to be fully understood. No consensus has been reached concerning
the phenomenology of these processes. An attractive possibility made use
of the intrinsic ambiguities of fixed order perturbation theory to
define the various unphysical scales entering the theoretical
predictions by means of various ``optimisation"
 prescriptions~\cite{GRUN,STEV,BLM,OPT}: an
excellent agreement~\cite{ABDFS,ABFOW,VV,GORD} between theory and
experiments over the whole available range of energy and transverse
momentum was thus obtained with a single set of structure functions and
a unique value of $\Lambda_{_{QCD}}$. More recently however it has been
proposed to fix the unphysical scales at some arbitrary ``physical"
values and to introduce an extra non-perturbative parameter, called the
``intrinsic transverse momentum" of the partons in the hadrons which is
fitted to the data at each energy~\cite{E706,HKKLOT,FERB,ABBHMT,ZIEL}.
This parameter increases with energy (technically speaking it is
therefore not an ``intrinsic" momentum) and this is interpreted as
taking into account the effects of multiple soft gluon emission
associated to the hard  partonic scattering.

Motivated by the recent publication of two new sets of fixed target data
and using the latest up-to-date theoretical calculations we discuss the
phenomenology of inclusive prompt photon production. The new data 
are those of the UA6 proton-proton and antiproton-proton experiment at a
center-of-mass energy $\sqrt s = 24.3$ GeV~\cite{UA6} and of the E706 proton-Beryllium
and pion-Beryllium experiment at $\sqrt s = 31.6$ GeV  and at $\sqrt s =
38.8$ GeV~\cite{E706}. To avoid further ambiguities associated to the criteria of
isolation we do not discuss the colliders data on the production of
isolated prompt photons.

We first set the theoretical framework by recalling the main features of
the complete next-to-leading order (NLO) calculations in perturbative Quantum
Chromodynamics (QCD). The intrinsic uncertainties of the NLO expressions
are related to the choice of three arbitrary scales: the renormalization
scale, the factorization scale associated to the initial state collinear
singularities and the fragmentation scale related to the the collinear
fragmentation of a parton into a photon. A rather complete numerical
study of these uncertainties is carried out. The main feature is that
there is no stability zone in the fixed target energy range unlike what
has been observed at higher energies~\cite{ACFGP}. Arbitrarily choosing
the scales we show that we can get a reasonable agreement with all considered
data except with the E706~\cite{E706} data obtained using nuclear
targets and the older R806 data~\cite{R806} from the ISR. This may raise
the possibility of inconsistency among various data sets.

\section{Theoretical framework and ambiguities}

Because of the well-known anomalous photon component the cross section
for inclusive photon production takes a more complicated form than for
pure hadronic reactions. The differential cross section in transverse
momentum $\pT$ and rapidity $\eta$ can be written as a sum of two
components:
\be
\dsig = \dsigd\ + \ \dsigb
\label{eq:sig}
\ee
where we have distinguished the ``direct'' component 
$\sigma^{\hbox{\cmr dir}}$ from the
``bremsstrahlung'' one $\sigma^{\hbox{\cmr brem}}$. Each of these terms is known in
the next-to-leading logarithmic approximation in QCD {\em i.e.} we have
\bea
\dsigd &=& \sum_{i,j=q,g} \int dx_{1} dx_{2}
\ F_{i/h_1}(x_{1},M)\ F_{j/h_2}(x_{2},M) \nonumber \\
&\ & \qquad \qquad
\alfspi \left( \dsigij \ + \ \alfspi \kd_{ij} (\mu,M,M_{_F}) \right)
\label{eq:dir}
\ena
and
\bea
\dsigb &=& \sum_{i,j,k=q,g} \int dx_{1} dx_{2}{ dz \over z^2}
\ F_{i/h_1}(x_{1},M)\ F_{j/h_2}(x_{2},M)\ D_{\gamma/k}(z, M_{_F})
\nonumber \\
&\ & \qquad \qquad
\ \left( \alfspi \right)^2\  \left( \dsigijk \ + 
\ \alfspi \kb_{ij,k} (\mu,M,M_{_F}) \right)
\label{eq:brem}
\ena 
where the parton densities in the initial hadrons $F_{i/h_1}$ and
$F_{j/h_2}$ depend on the factorization scale $M$ while the parton to
photon fragmentation functions $D_{\gamma/k}$ depend on the fragmentation
scale $M_{_F}$. The renormalization scale $\mu$ appears in the strong
coupling $\alpha_s$. The higher order correction terms to the direct and
bremsstrahlung cross sections, $\kd_{ij}$~\cite{ABDFS,GV} and
$\kb_{ij,k}$~\cite{ACGG} respectively, are known and we shall use their
expressions in the $\MSB$ convention. The dependence of these functions
on the kinematical variables $x_1, x_2, z, \sqrt s, \pT$ and $\eta$ has
not been explicitly displayed. All the quantities entering the above
equations have been either calculated ($\kd_{ij}$ and $\kb_{ij,k}$) or
have been determined  (see $e.g.$ \cite{ABFOW,CTEQ,MRS,GRVdis} for $F_{i/h}$
and \cite{GRV,BFG} for $D_{\gamma/k}$) at the required level of accuracy
by next-to-leading order fits to the data. The knowledge of $\LMS$, from
deep-inelastic experiments for example,  completely specifies the NLO
expression of the running coupling $\as$. It is important to stress that
the scales $\mu$, $M$ and $M_{_F}$ are arbitrary. The dependence in the
first two scales partially compensates within each of eq. (\ref{eq:dir})
and eq.~(\ref{eq:brem}) between the lowest order term and the correction
term. 
The dependence in the fragmentation scale is more complex because of the
dual nature of the photon which acts either as a parton (anomalous part)
or a hadron. We come back to this point in great detail below. For our
present purposes, it is enough to know that the $M_{_F}$ dependence of
the photon fragmentation function is compensated partly by terms in
$\kd$ of eq.~(\ref{eq:dir}) and partly by terms in $\kb$ of eq.
(\ref{eq:brem}). Since the scale $M_{_F}$ is arbitrary, it is clear that
eq.~(\ref{eq:dir}) and eq.~(\ref{eq:brem}) which depend essentially
monotonically on the scale $M_{_F}$ do not have an independent physical
meaning: only the sum, eq.~(\ref{eq:sig}), has a chance to be
(relatively) independent of $M_{_F}$ and therefore is interpreted as a
physical quantity. 

%
\begin{figure}[htb]
\vskip 1.cm
\begin{picture}(484,227)(0,0)
\put(-21,0){\epsfig{file=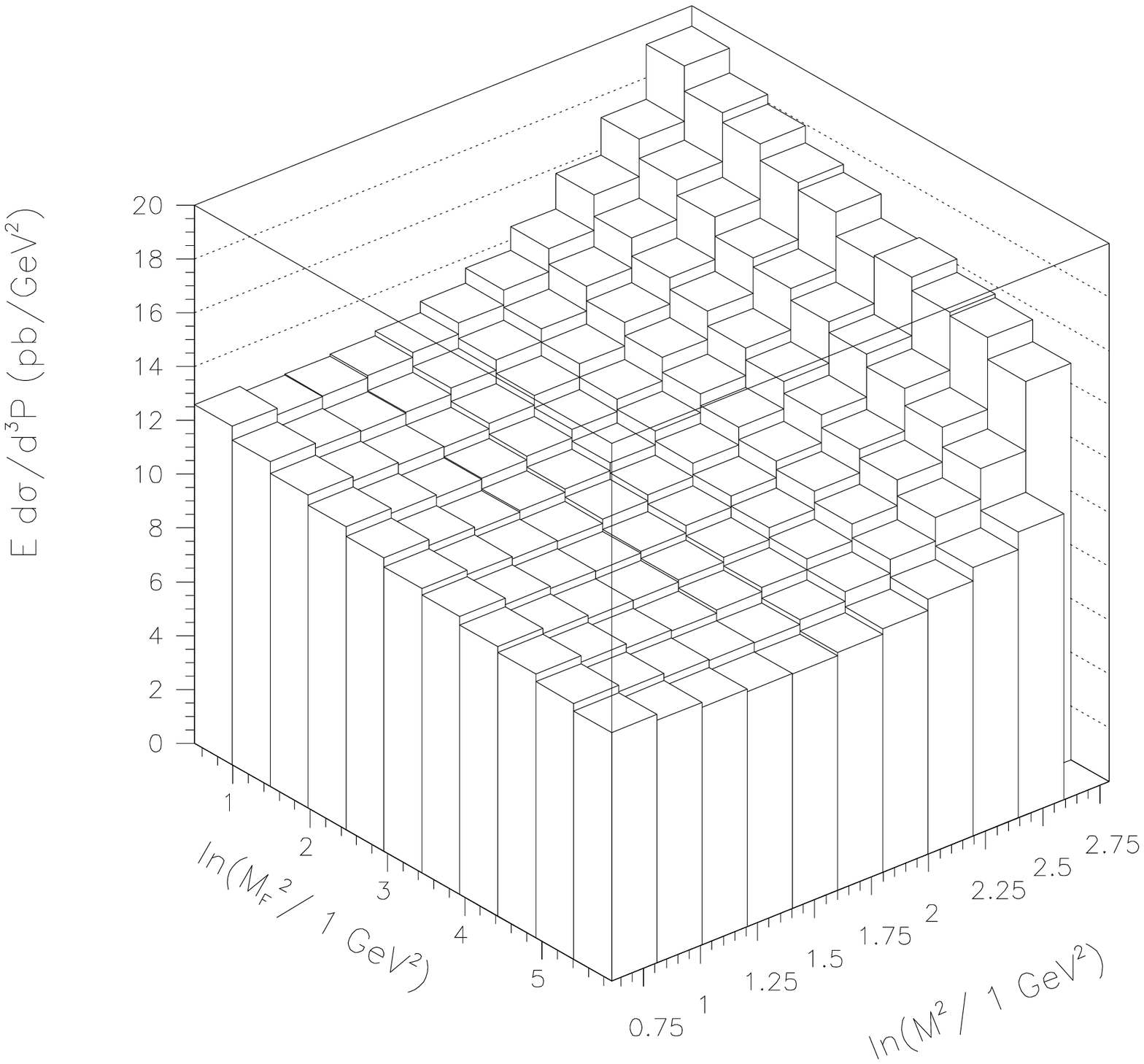,width=10.cm}}
\put(210,0){\epsfig{file=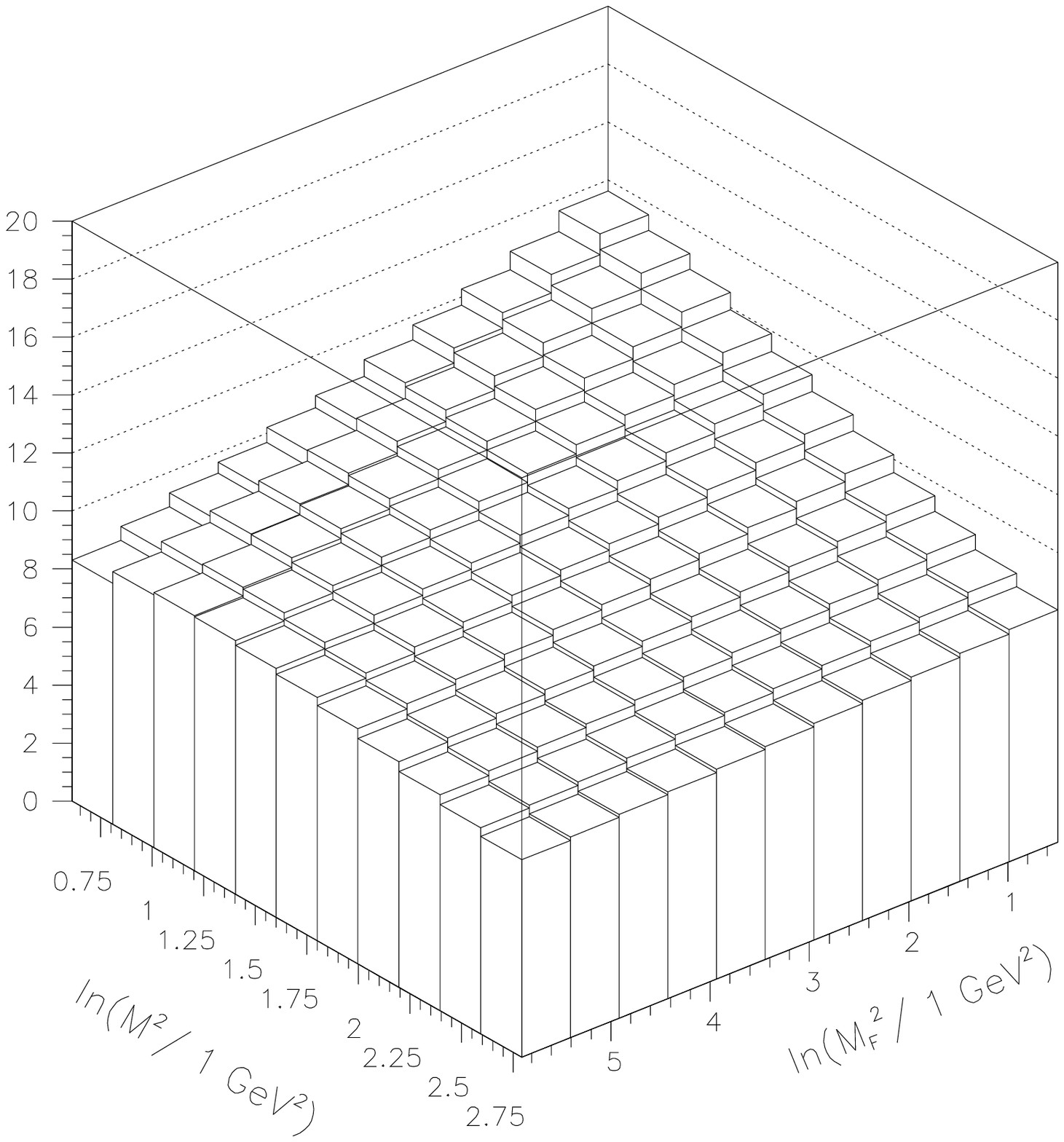,width=10.cm}}
\end{picture}
\vskip -.5cm
\caption{\em
Variation of the inclusive photon production cross section for
proton-beryllium at $E_{\hbox {lab}} = 800$ GeV, $p_{_T} = 6.12$ GeV/c,
$\eta = 0.$, as a function of the factorization scale and the
fragmentation scale; {\em left}: the renormalization scale is
``optimized" according to  eq. (\ref{eq:opt}); {\em right}: the
renormalization scale is set equal to the factorization scale.
}
\label{secmmf}
\end{figure}
The fundamental ambiguity of the perturbative prediction is due to the
fact that the three scales $\mu$, $M$ and $M_{_F}$  are not determined
by the theory. Roughly speaking they are parameters which control how
much of the higher order effects are resummed in $\as$, $F_{i/h_1}$ and
$D_{\gamma/k}$ respectively and how much is treated perturbatively  in
$\kd_{ij}$ and $\kb_{ij,k}$.
In order to illustrate this problem keeping however the
numerical work at a reasonable level we proceed as follows. We fix
arbitrarily the fragmentation scale $M_{_F}$ and the factorization scale
$M$. The remaining scale $\mu$ is then determined by solving the
algebraic equation \cite{STEV,OPT}
\bea
{d \left( d \sigma  / d {\vec p}_{_T} d \eta \right) 
	\over d \ln(\mu) } \ = \ 0.
\label{eq:opt}
\ena   
The result is illustrated in the left picture of fig.~\ref{secmmf} for
the inclusive production of photons in proton-beryllium reaction at $\rs
= 38.8$ GeV, $\pT=6.12$ GeV/c. We see that at fixed $M_{_F}$ there is
a stability point (minimum) of the cross section as $M$ varies which is
a feature of a reliable perturbative prediction. However this stability
point does not go through an extremum when the fragmentation scale
varies.  An alternative choice for the renormalization scale $\mu$
consists in setting, for example, $\mu=M$ rather than using
eq.~(\ref{eq:opt}). This leads to the right hand picture in
fig.~\ref{secmmf}: again no stability is achieved in the fragmentation
scale while a stable point (maximum) appears when varying $\mu=M$ at
fixed $M^2_{_F}> 24$~GeV$^2$.  Therefore, we do not have a stable point on
the $M$, $M_{_F}$ sheet and one is somewhat at a loss about which scale
one should choose to make predictions. This pessimism should be tempered
however by noticing that there is a wide domain of relative stability
for the values $M^2 < 10$ GeV$^2$ and $M^2_{_F} < 300$ GeV$^2$ (left
picture) or  $M^2_{_F} > 24$ GeV$^2$ where a maximum in $M$ can be
obtained (right hand picture). Despite the completely different scale
variation patterns of the two pictures in fig.~\ref{secmmf} it should be
noted that, for a given fragmentation scale, the theoretical cross
section  estimates at the stable points are numerically very close: they
differ by less than $1\%$. Finally, when increasing $M_{_F}^2$ from 24
GeV$^2$ to 300 GeV$^2$ the optimal predictions decrease by less than
$20 \%$.  
It should be reminded that at much higher energies, in the Tevatron
energy range and above, a stable point could be obtained for the
inclusive photon cross section \cite{ACFGP}. 

\begin{figure}[htb]
\begin{center}
\vskip -1.cm
\epsfig{file=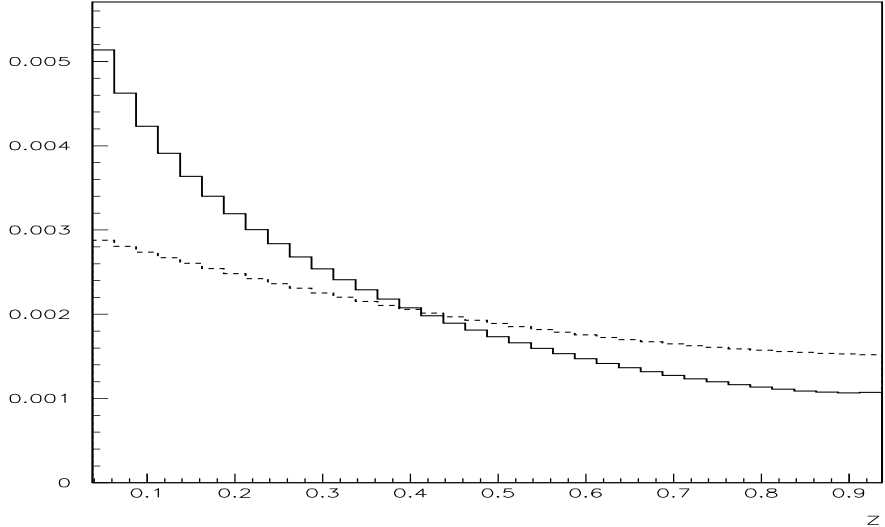,height=8.cm,width=14.cm}
\end{center}
\vskip -1.cm
\caption{\em 
{\em solid line}: $z$ dependence of the perturbative component of the
photon fragmentation function~eq.~(\ref{eq:approx});  
{\em dashed line}:  $z$ dependence of the compensating term
proportional to ${\bf P}^{(0)}_{\gamma q}(z))$. The scale is
$M_{_F}=6$~GeV.
}
\label{zplot}
\end{figure}

To understand the reason of the present unstability under variation of
the fragmentation scale one can make the following observations.
Consider the evolution equation of the photon fragmentation function. It is
given by (dropping the longitudinal variable $z$ and denoting instead the
convolution in $z$ by the notation $\otimes$)
\bea
{d D_{\gamma/i} (M_{_F}) \over d \ln (M_{_F}^2)}\ =\
{\bf P}_{\gamma i} \ +\ \asmfpi \sum_{j=q,g} P_{j i} 
\otimes D_{\gamma/j}
\label{eq:evol}
\ena 
where the ${\bf P}_{\gamma i}$ and $P_{j i}$ are the appropriate
Altarelli-Parisi splitting functions which admit a perturbative
expansion in $\alpha_s$. For consistency we keep the first two terms in
our analysis, namely
\bea
{\bf P}_{\gamma i} \ =\ {\bf P}^{(0)}_{\gamma i}
	\ + \ \asmfpi \ {\bf P}^{(1)}_{\gamma i}
\label{eq:anoma}
\ena
and similarly for $P_{j i}$.
The solution to eq.~(\ref{eq:evol}) can be written as the sum
of the anomalous part and the non-perturbative part:
\bea
D_{\gamma/k} (z,M_{_F}) = D^{^{AN}}_{\gamma/k} (z,M_{_F}) + 
D^{^{NP}}_{\gamma/k} (z,M_{_F}).
\label{eq:frag}
\ena
The component $D^{^{AN}}_{\gamma/k}(z,M_{_F})$ is fully calculable and
it is a particular solution of the full inhomogeneous set of eq.
(\ref{eq:evol}). 
The non-perturbative part (modeled following the Vector Meson
Dominance (VMD) ideas) $D^{^{NP}}_{\gamma/k} (z,M_{_F})$ obeys the usual
homogeneous evolution equations of hadronic structure functions  {\em
i.e.} eq.~(\ref{eq:evol}) with ${\bf P}_{\gamma i}=0$. The $M_{_F}$
scale dependence of $D^{^{NP}}_{\gamma/k}$, appearing in $d
\sigma^{\hbox{\cmr brem}}$, compensates between the two terms of the
right hand side of eq.~(\ref{eq:brem}) as it should be for purely
hadronic reactions. In contrast, the scale variation of
$D^{^{AN}}_{\gamma/k}$ associated to the inhomogeneous term ${\bf
P}_{\gamma k}$ is compensated by a similar term in $\kd_{ij}$ of
eq.~(\ref{eq:dir}). It turns out that, in the kinematical region probed
by fixed target experiments, such a scale compensation is not properly
achieved. Indeed, due to steeply falling partonic cross sections,
the average value of the variable $z$ in the bremsstrahlung process is
rather high, $z \sim .8$. In that domain, we can safely neglect the
gluon fragmentation function (as well as the non-perturbative VMD input)
and check that the approximation
\bea
D_{\gamma/q}(z,M_{_F})&=&{ {\bf c}_{\gamma q}(z) \over \alpha_s(M_{_F}) } 
			+ {\bf d}_{\gamma q}(z) \nonumber \\
&\sim& b\ {\bf c}_{\gamma q}(z) \ln(M_{_F}^2/\LMS^2)
			+ {\bf d}_{\gamma q}(z),
\label{eq:approx}
\ena
with the moments (${\bf c}(n) = \int^1_0 dz z^{n-1}{\bf c}(z)$) of the
${\bf c}_{\gamma q}(z)$ function given by
\bea
{\bf c}_{\gamma q}(n) = { {\bf P}^{(0)}_{\gamma q}(n) 
				\over b - P^{(0)}_{q q}(n)/2 \pi },  
\label{eq:moment}
\ena
satisfy eq.~(\ref{eq:evol}) (neglecting ${\cal O}(\alpha_s)$ terms).
In the above equations the parameter $b$ is the well known positive
constant governing the evolution of the strong coupling in the leading
logarithmic approximation
\bea
{d \alpha_s(M_{_F}) \over d \ln (M^2_{_F})} = 
- b\ \left( \alpha_s(M_{_F}) \right)^2\ (1 + b_1 \alpha_s(M_{_F})).
\label{eq:coup}
\ena
At large $z$, the function  $ b\ {\bf c}_{\gamma q}(z)$ is much smaller
than the function ${\bf P}^{(0)}_{\gamma q}(z)$ which appears as the
coefficient of the $\ln (M^2_{_F})$ dependence in eq.~(\ref{eq:dir}).
This can be guessed from eq.~(\ref{eq:moment}) since $P^{(0)}_{q q}(n)$
is negative at large $n$, equivalently large $z$, as required by the
negative  scaling violations of the quark fragmentation function. The
relative smallness of $b\ {\bf c}_{\gamma q}(z)$ is understood by the
fact that it represents the shape of the fragmentation function of the
quark into a photon after multiple gluon emission while  ${\bf
P}^{(0)}_{\gamma q} (z)\sim (1+(1-z)^2)/z$ describes the fragmentation  at
the lowest order level (without gluon emission).  The overall scale
$M_{_F}$ dependence of eq. (\ref{eq:sig}) is then a weighted average
over large $z$ values of the combination
\bea 
( b\ {\bf c}_{\gamma q}(z) - {\bf P}^{(0)}_{\gamma q}(z)) \ln (M_{_F})
\label{eq:scavar}
\ena
which is always negative in the large $z$ range, in agreement with the
results of fig.~\ref{secmmf}.  
This point is illustrated more precisely on fig.~\ref{zplot} which compares,
at a fixed value of $M_{_F}$, the variation in $z$ of $D_{\gamma/q}$
of eq.~\ref{eq:approx} with that of the compensating term proportional to
${\bf P}_{\gamma q}$. As anticipated, the compensating term has a less
steep dependence and largely dominates at large $z$. Since the effective
value of $z$ in inclusive photon production at E706 energies increases
from $z \sim .85$ to about $z \sim .93$ when $\pT$ varies from 5 GeV/c
to 10 GeV/c it is clear that  expression (\ref{eq:scavar}) is negative
in the region of interest{\footnote{As $\pT$ increases the relative
importance of the fragmentation terms decreases compared to direct 
photon production terms.}}.

This discussion concentrated only on the scale
variation driven by the inhomogeneous term (${\bf P}_{\gamma q}$) in
eq.~(\ref{eq:evol}). The  homogeneous term, included in our calculation,
somewhat weakens the dependence in expression (\ref{eq:scavar}) but does
not change the overall behaviour. We do not see how to overcome this
basic instability of the perturbative prediction for the inclusive
production of direct photon in hadronic collisions. 
Notice that the perturbative approach is not spoiled by the presence of
large $\ln (1-z)$ terms, for they cancel each other (at order $\alpha_s^2$)
between the direct contribution eq.~(\ref{eq:dir}) and the
bremsstrahlung contribution eq.~(\ref{eq:brem})~\cite{BFG}.
Recently, two groups~\cite{CMN} have studied, in hadronic processes, the
resummation of collinear and soft logarithms which become large in the
limit where $x_1$ and $x_2$ are close to 1. Although the average values
of $x_1$ and $x_2$ in eqs.~(\ref{eq:dir},\ref{eq:brem}) are of the
order of $x_{_T}$, the resummed expressions might slightly change the
values of the predictions obtained in this paper. The stability with
respect to the scales $\mu$ and $M$ could also be improved. These types
of corrections are not considered here. It would be interesting to
investigate their impact on the prompt photon phenomenology.

We propose to bracket the theoretical results by chosing two extreme
values of the fragmentation scale. Let us note that making the choice
$M_{_F} = \mu$ may accidentally reduce the overall scale dependence of
the cross section since, for example, one power of $\as$ in
eq.~(\ref{eq:brem}) is compensated by  the $1 / \as$ factor of
$D_{\gamma/q}$. We do not advocate this choice although we will often
use it for practical reasons in the following.

Before closing these theoretical preliminaries, let us make a technical
comment. It concerns the definition of $\as$. The solution of the NLO
evolution equation of the coupling (eq.~(\ref{eq:coup})) leads to the
following relation in the $\MSB$ scheme
\bea
b \ \ln (\mu^2 / \LMS^2) = {1 \over \alpha_s(\mu)} + b_1
 \ln \left( {b \ \as \over 1+ b_1 \as} \right).
\label{eq:alfas}
\ena
where $b=(33-2N_f)/12 \pi$ and $b_1=(153-192N_f)/( 2 \pi(33-2N_f))$ (with
$N_f=4$, the number of flavors).
To obtain $\as$ explicitly one may invert exactly (numerically) this
equation, or one may invert it approximately (analytically) keeping the
first two terms in the expansion of $\as$ in terms of $1/ \ln(\mu^2 /
\LMS^2)$. In the region of interest for this study the difference
between the obtained values of $\as$ for a given  $\mu$ may reach $5\%$.
We advocate, and we will use in the following, the ``exact"
estimate of $\as$.

\section{Phenomenology of prompt photon production}

Let us first summarize the data sets used in the following discussion.
This is done in table~\ref{data}. To avoid uncertainties due to the
pion structure functions we restrict our discussion to the case of
proton induced reactions.
Concerning E706 and the possible nuclear effects on the parton
distributions in the Beryllium target we follow the same prescription as
the collaboration and adjust the theoretical predictions with the factor
$A^{\alpha-1}$ with $\alpha=1.04$. In this way the resulting cross
section is normalized per nucleon. 
\begin{table}[t]
\vskip -.5cm
\begin{center}
  \begin{tabular}{|c|c|c|c|c|c|}
    \hline 
    Collaboration & Reaction & $\rs$ & $p_{_T}$ range & $x_{_F}$/rapidity & $x_{_T}$ range \\
     & & [GeV] & [GeV/$c$] & range  & \\
 \hline \hline 
    WA70\cite{WA70} & $ p\ p$ & 23.0 &  $p_{_T}> 4.0$ & $-.35 < x_{_F} <.45$  & $x_{_T}> .35 $\\
 \hline
    UA6\cite{UA6}  & $ p\ p$ & 24.3 &  $p_{_T}> 4.1$ & $-0.1 < \eta <0.9$  & $x_{_T}> .34 $ \\
 \hline
    UA6\cite{UA6} & $ \bar p\ p$ & 24.3 &  $p_{_T}> 4.1$ & $-0.1 < \eta <0.9$ &  $x_{_T}> .34 $\\
 \hline
    E706\cite{E706} & $ p\ Be$ & 31.6 &  $p_{_T}> 3.5$ & $-.75 < \eta <.75$ &  $x_{_T}> .22 $\\
\hline
    E706\cite{E706} & $ p\ Be$ & 38.8 &  $p_{_T}> 3.5$ & $-1.0 < \eta <0.5$ &  $x_{_T}> .18$ \\
\hline
    R806\cite{R806} & $ p\ p$ & 63. &  $p_{_T}> 3.5$ & $ -0.2 <\eta <0.2$ &  $x_{_T}> .11 $\\
\hline
    R110\cite{R110} & $ p\ p$ & 63. &  $p_{_T}> 4.5$ & $ -0.8 <\eta <0.8$ &  $x_{_T}> .14 $\\
\hline
AFS/R807\cite{R807} & $ p\ p$ & 63. &  $p_{_T}> 4.5$ & $ -0.7 <\eta < 0.7$ &   $x_{_T}> .14 $\\
\hline
  \end{tabular}
\end{center}
\caption{\em 
Summary of the main features of the data sets. The transverse and
longitudinal variables $x_{_T}$, $x_{_F}$ are defined by $x_{_T}=
2\pT/\sqrt s$ and $x_{_F}=2 p_{_L}/ \sqrt s$.
}
\label{data}
\end{table}
\begin{figure}[ht]
\vspace{-2.cm}
\begin{center}
\epsfig{file=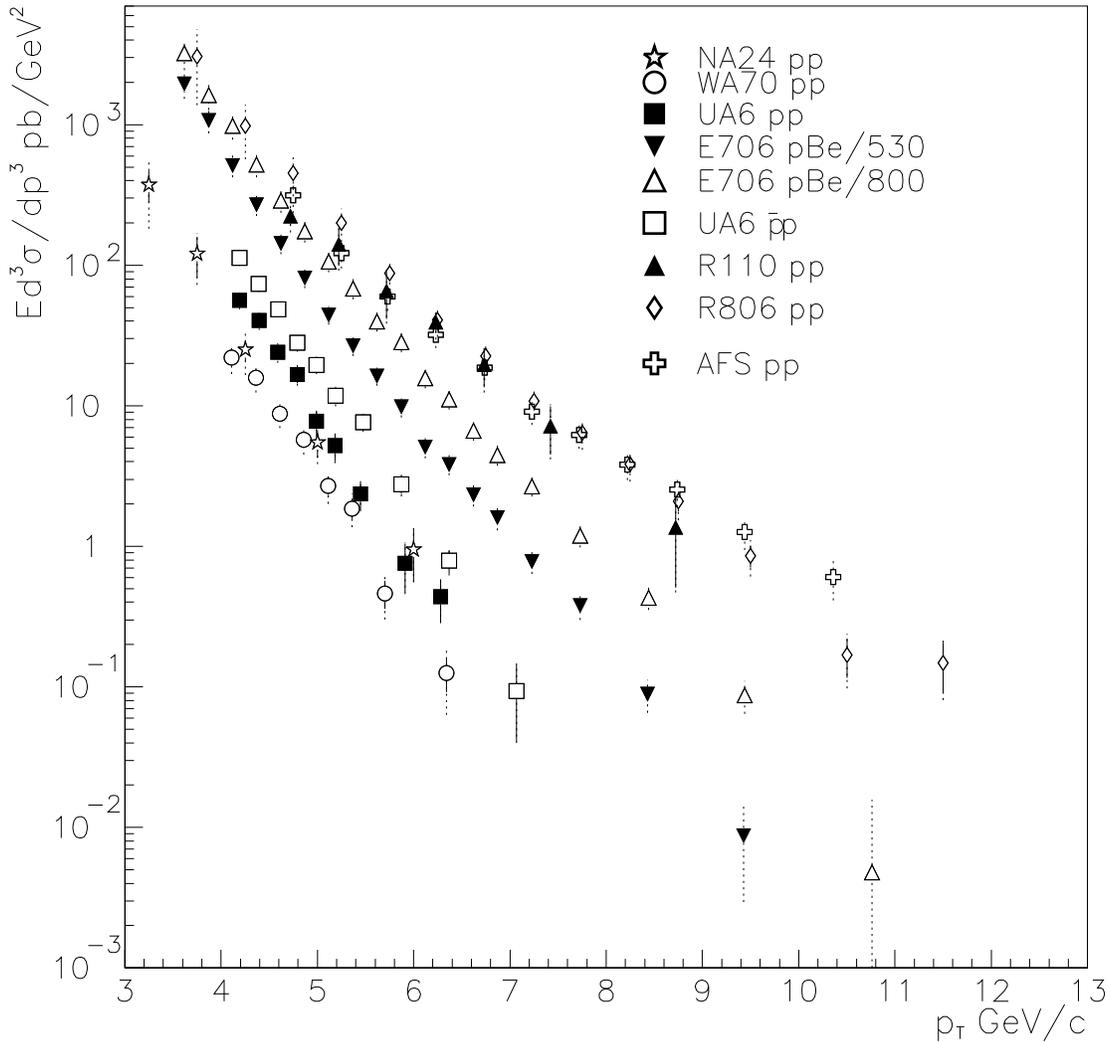,height=16cm}
\end{center}
\vskip -1.cm
\caption{\em 
The differential inclusive photon cross sections as a function of
transverse momentum. The cross sections are averaged over the
$x_{_F}$/rapidity ranges shown in table~\ref{data}. Data from the NA24
collaboration \cite{EXP} are averaged in the rapidity range $-.65 < \eta
< .52$.
}
\label{dataset}
\end{figure}
The inclusive differential cross sections are shown in fig.~\ref{data}
where they  are plotted as a function of $\pT$.   Already, we anticipate
some difficulties when comparing the data sets: for example, the
relatively large difference between the WA70 and UA6 cross sections at
low $\pT$ cannot be accounted for by the small $\sqrt s$ change between
experiments nor by the different rapidity coverage; one also observes
almost an order of magnitude difference between the UA6($p p$) data and
the E706(530 GeV) for an energy change from $\sqrt s = 24$ GeV to $\sqrt
s = 31.6$ GeV while the E706(800 GeV) and R806 data at low $\pT$ have
almost the same size despite an energy increase from $\sqrt s = 38.8$
GeV to $\sqrt s = 63.$ GeV.

In the numerical calculations we will use several sets of recent
next-to-leading order structure functions in the $\MSB$ convention.
CTEQ~\cite{CTEQ} and MRS~\cite{MRS} will be our main sets. However, in
order to compare the present results with the ``old" fits done using the
WA70 data we will also make use of the original ABFOW~\cite{ABFOW} sets
{\footnote {In our previous phenomenological studies 
	\cite{ABDFS,ABFOW} the higher order calculations
  	to the bremsstrahlung cross section were not available and 
	a stability in the fragmentation scale could not be expected. 
	That scale was then chosen to be $\hat s$.}}.      
Concerning the higher order
calculations they have been performed using four light flavors and therefore
$m_{\hbox {\cmr charm}}=0$ in the partonic cross sections and the higher order
terms. However the threshold effects due to the charm quark mass are taken into
account in the fragmentation function $D_{\gamma/c}$ in such a way that at a
scale $M_{_F} \sim m_{\hbox {\cmr charm}}$ one recovers all the correct
logarithmic factors of the exact massive calculation~\cite{BFG}. The most recent
parametrization of the NLO parton to photon fragmentation functions of
ref.~\cite{BFG} will be used. In this work, two parametrizations of the
fragmentation functions $D_{\gamma/i}$, based on fits to $e^+ e^-$ data,
are given: we use set II parametrization which corresponds to the large
non-perturbative gluon input. In fact, little difference is observed in
inclusive photon production in hadronic collisions between set I and set
II predictions since, as discussed in the previous section, the
effective fragmentation variable $z$ is large and the fragmentation
functions are then dominated by their perturbative components.

We first discuss the comparison between theory and experiments for
proton-proton and proton-antiproton (when applicable) collisions, using
a fixed common scale $\mu=M=M_{_F}$. Then we shall  relax this
constraint and study optimised predictions in the $\mu$ and $M$ scales
for a range of fragmentation scale in order to propose an estimate of
the errors on the perturbative predictions. In the theoretical
predictions we always include the NLO corrections to both the direct
component and to the bremsstrahlung component (denoted ACFGP in the
figures).

\begin{figure}[htb]
\begin{center} 
\vskip -1.cm
\epsfig{file=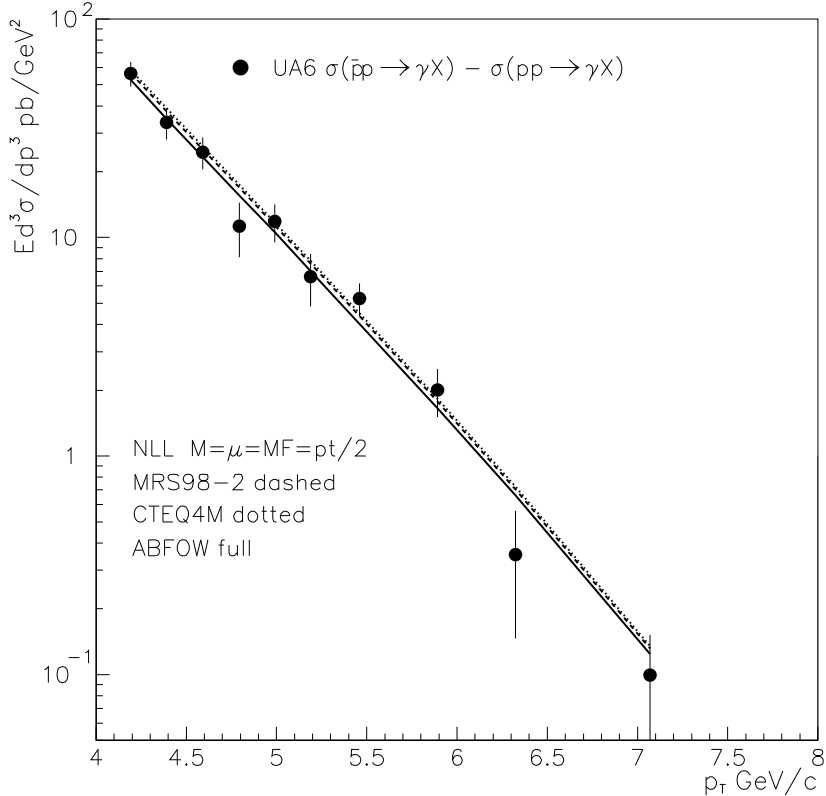,height=12cm}
\end{center}
\vskip -1.cm
\caption{\em 
Comparison of the UA6 $\bar p p- p p$ data~\protect\cite{UA6}
with the NLO theory for various sets of structure functions
using the common scale $\mu=M=M_{_F}=\pT/2$. Only statistical errors
are shown.
}
\label{difua6}
\end{figure}

In order to normalize our present results to the previous ones
appropriate to WA70 data we start our analysis using the ABFOW
parametrization ($\LMS = 230$ MeV, $x\ G(x,Q_0^2=2\ {\mbox{\rm GeV}}^2)
= (1-x)^4$). We get with all scales in the problem set equal to $\pT/2$
a good approximation of the old optimised results and also  a good
agreement with the WA70 data  ($\chi^2 = 8.2$ for 7 points using
statistical errors only{\footnote{As in \cite{ABFOW}, the lowest $p_T$
point is not included in the $\chi^2$ evaluation, due to important
systematic errors
.}}). Comparing with the new UA6 measurements, there is an excellent
agreement with the data (see fig.~\ref{difua6}) on the
difference $\bar p p - p p$ (which is insensitive to the gluon density
in the proton and to the parton fragmentation functions into a photon)
while the $\bar p p$ data tend to be higher than
\begin{figure}[htb]
\vskip -1.cm
\begin{center}
\epsfig{file=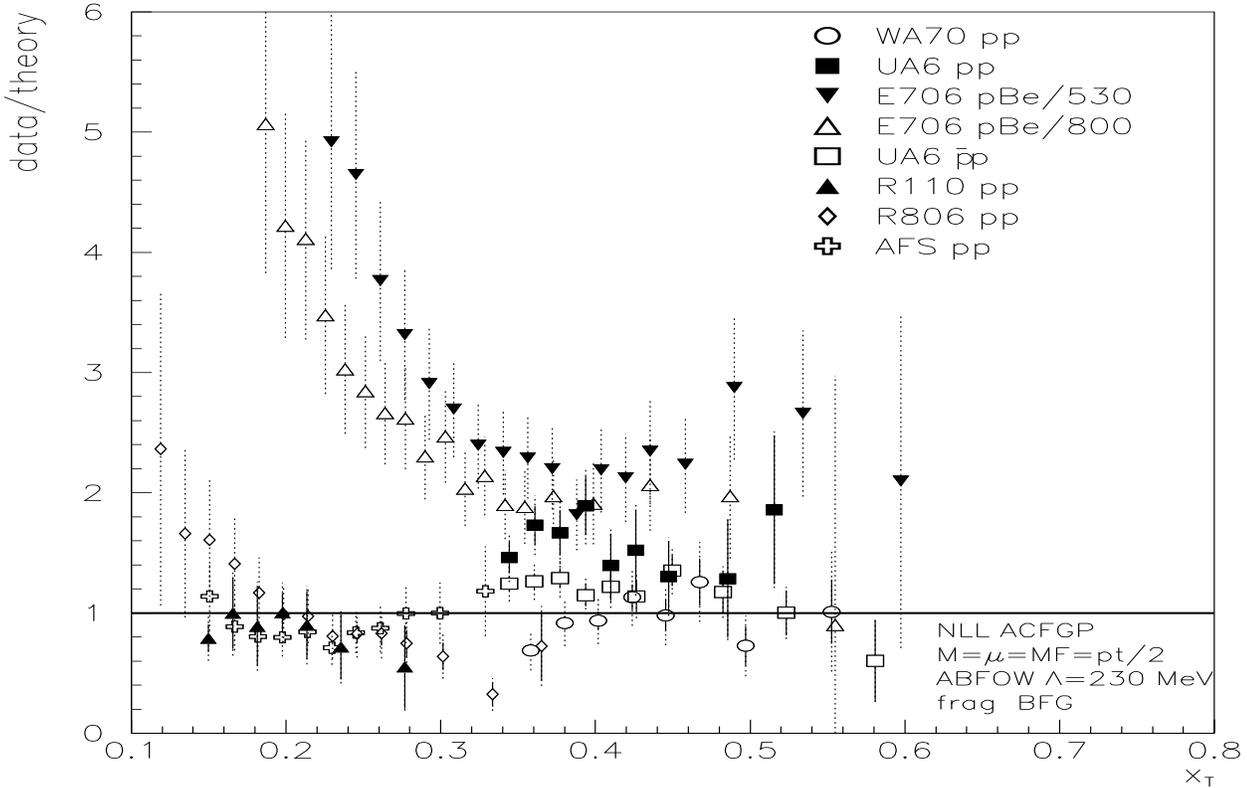,height=12cm,width=18.cm} 
\end{center}
\vskip -1.cm
\caption{\em
Comparison of data with theory, normalized to the theoretical results,
for the various experiments as a function of $x_{_T}$. The ABFOW
structure functions are used and all scales are set equal to $\pT/2$.
Statistical error bars are shown as full lines while systematic
uncertainties are added in quadrature and shown as dotted lines. 
An extra 5\% uncertainty on the energy scale is not plotted for R110;
for AFS and E706 data, statistical and systematic errors are shown
combined in quadrature only.  
} 
\label{rats2}
\end{figure}
the theory by roughly $20\%$ and the $p p$ data by almost a factor 2 for
$x_{_T}<0.4$ ($\chi^2 = 64.4/9$).  The normalized ratios $data/theory$
are shown in fig.~\ref{rats2}. At large $x_{_T}$ the WA70 and UA6 data
are mutually compatible within the error bars while there may be some
relative normalization problem for the points below $x_{_T}=0.4$.
Turning now to the ISR results one notices that they are located at
small $x_{_T}$ values and that, for the R110 and R807 data, the ratio
$data/theory$ is flat as a function of $x_{_T}$; furthermore it is
compatible with 1 ($\chi^2 = 6.8/7$ for R110 and $\chi^2 = 13.7/11$ for
R807{\footnote {All the above $\chi^2$ values are obtained using
statistical errors only except in the case of AFS/R807 where only the
sum of statistical errors and systematic errors added in quadrature is
available.}}). In contrast, the R806 data show a marked decrease from
1.5 to 0.5 as $x_{_T}$ increases. It is clear that the problem of the
R806 data is not related so much to the overall normalization as to the
slope of the $x_{_T}$ distribution. In fact, it was noted before that
the $\pT$ dependence of the R806 data was incompatible with the  slope
of the other direct photon experiments. Besides, it was not  possible to
obtain a good fit simultaneously to the deep-inelastic data and to the
R806 data~\cite{ABFOW}. However all ISR sets, up to $x_{_T} = 0.3$ are
point to point compatible taking into account statistical as well as
systematic error bars. 

The main surprise comes when comparing the E706 data to theory: at
both energies we observe that the experimental points are 2 to 3 times
larger than the theoretical predictions at high $x_{_T}$ and the ratio
experiment/theory keeps increasing up to a value of 5 as $x_{_T}$
decreases. This had already been stressed by the E706 
collaboration~\cite{E706}. 

\begin{figure}[hbt]
\vskip -1.cm
\begin{center}
\epsfig{file=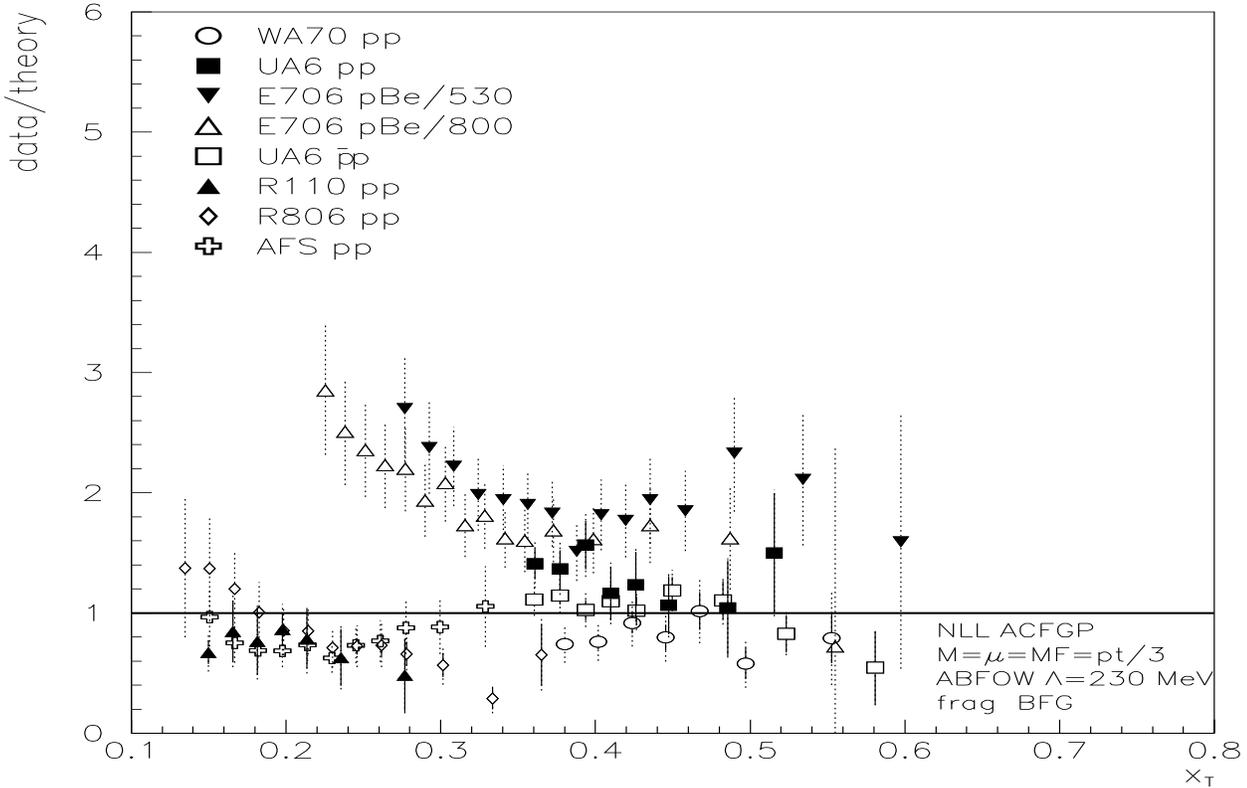,height=12cm,width=18.cm}
\end{center}
\vskip -1.cm
\caption{\em
Comparison of data with theory, normalized to the theoretical results,
for the various experiments as a function of $x_{_T}$. The ABFOW
structure functions are used and all scales are set equal to $\pT/3$. 
Data points at scales such that $(\pT/3)^2 < Q_0^2=2.$ GeV$^2$ are not
included, which explains why the points at lower $x_{_T}$ values in some
experiments are not displayed.  
See fig. {\protect\ref{rats2}} for further comments.
}
\label{rats3}
\end{figure}

A similar study can be conducted with the common scale in the
calculation set equal to $\pT/3$ instead of $\pT/2$. The corresponding
results are displayed in fig.~\ref{rats3}.
Good agreement is achieved for the difference $\bar p p - p
p$ and the UA6 $\bar p p$ data ($\chi^2 = 14.7/9$). Agreement between
theory and experiment is much improved in the case  the UA6 $p p$ data
($\chi^2 = 23.1/8$) while, on the contrary, the comparison between WA70,
R110 and R807 is not so satisfactory (all $\chi^2$ values are roughly
multiplied by a factor 3 compared to the analysis with scales $\pT/2$).
In fact, all these data sets fall somewhat below the theoretical
predictions.
Concerning E706, essentially the same situation as before
prevails since the theoretical predictions are rather stable (they
increase by about 20 \%) when changing the common scale from $\pT/2$ to
$\pT/3$, namely the theory still underestimates the data by roughly a
factor 2 at large $x_{_T}$ and much more at small $x_{_T}$.\\ 

\begin{figure}[htb]
\vskip -1.cm
\begin{center}
\epsfig{file=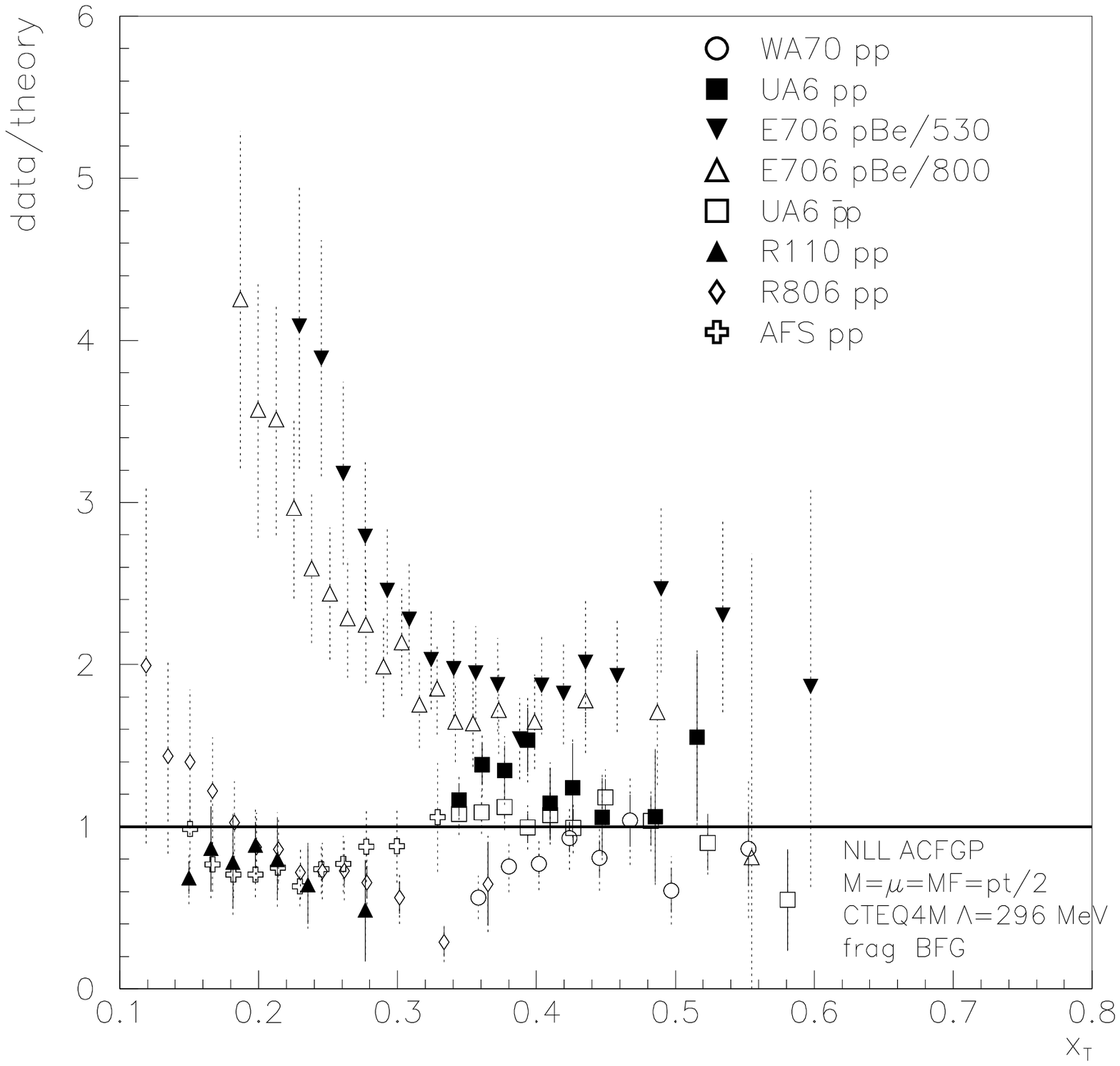,height=12cm,width=18.cm} 
\vskip -1.cm
\caption{\em
Dependence in $x_{_T}$ of the experimental cross sections normalized to
theoretical predictions using the CTEQ4M distributions. All scales in
the calculations have been set to $\pT/2$ and only data points
satisfying the condition $(\pT/2)^2 > Q_0^2=2.56$ GeV$^2$ are kept.
See fig. {\protect\ref{rats2}} for further comments.
} 
\label{xtcteqpt2}
\end{center}
\end{figure}
\begin{figure}[htb]
\vskip -1.cm
\begin{center}
\epsfig{file=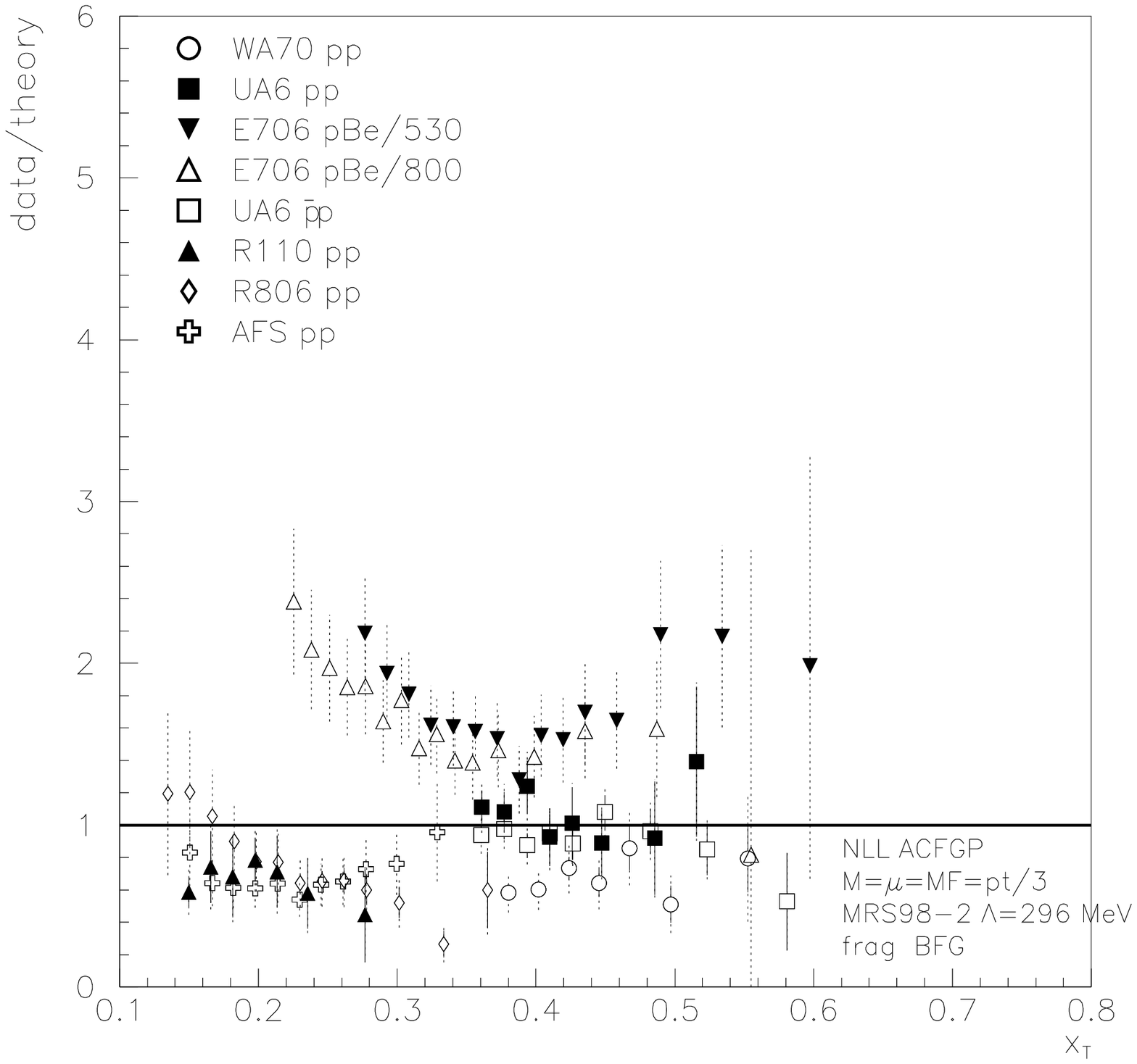,height=12cm,width=18.cm} 
\vskip -1.cm
\caption{\em
Dependence in $x_{_T}$ of the experimental cross sections normalized to
theoretical predictions using the MRS-98-2 distributions. All scales in
the calculation have been set to $\pT/3$ and only data points satisfying
$(\pT/3)^2 > Q_0^2=2.$ GeV$^2$ are shown.
The condition $\pT > 5.$ GeV/c  (see sec.~\ref{pheno})
translates into $x_{_T} > .32$ for E706~(530~GeV) and $x_{_T} >.26$ for
E706~(800~GeV) and $x_{_T} > .16$  for ISR data.
See fig. {\protect\ref{rats2}} for further comments.
}
\label{xtmrs982}
\end{center}
\end{figure}

The provisional conclusion to be drawn is that there is room for a minor
incompatibility in the normalization of the UA6 and WA70 $p p$ data at
low $\pT$ while they are in perfect agreement at large transverse
momentum, or to put it differently, there may be some incompatibility
in the $\pT$ dependence of the two sets of data. The ISR data are also
compatible with WA70/UA6 while the E706 results stand much above all
other experiments. It is to be noted that the UA6 $\bar p p$ data
are in good agreement with the theoretical predictions both
in shape and normalization.

We repeat our numerical exercises using the CTEQ4A1, CTEQ4M and MRS-98
structure functions. In the case of CTEQ4A1 we obtain very much the same
phenomenology as before. As illustrations, we show the comparison
between data and theory using CTEQ4M with all scales equal to $\pT/2$
in fig.~\ref{xtcteqpt2}{\footnote{The difference on the 
ratios $data/theory$ observed for the  E706 points, at low $\pT$,
between fig.~{\protect{\ref{xtcteqpt2}}} and fig.~14 of 
ref.{\protect{~\cite{E706}}} 
arises partly because in the latter work the higher order terms in 
eq.~{\protect{({\ref{eq:brem}}})
were not taken into account.}} and using MRS-98-2 with all scales equal to
$\pT/3$ in fig.~\ref{xtmrs982}. Fig.~\ref{xtcteqpt2} looks very similar
to the case of ABFOW with the smaller common scale $\pT/3$: this is
understood because of the larger value of $\Lambda_{_{\overline {MS}}}=
296$ MeV for CTEQ4M (instead of $230$ MeV) which yields a larger value
of $\alpha_s$ for a given scale and therefore larger theoretical
predictions. The same comment would apply to the results based on
MRS-98-2 with the common scale $\pT/2$. From fig.~\ref{xtmrs982}, there
appears a perfect agreement between theory and the UA6 $p p$ and $\bar p
p$ data (the $\chi^2$ values are 3.12/8  and 5.5/9 respectively)  and a
tendency to reduce the disagreement with  E706 to an overall
normalization factor of roughly 50 $\%$ at the cost of overestimating
the ISR data and the low $\pT$ WA70 data. Let us remark that the overall
improved agreement between theory and experiment is, in part, due to
dropping a few experimental points at low $\pT$ because of the scale
limitation.


It appears difficult to reconcile the E706 data with the other fixed
target data: clearly their normalization is higher. At large
$x_{_T}$ values ($x_{_T}> .3$), one can roughly estimate from the
previous figures that, when normalized to any perturbative NLO
predictions, they are above the other data sets by a factor ranging from
at least 50\% (compared to UA6, all scales $\pT/3$) to 2.0 (compared to 
WA70, all scales $\pT/2$). 

A further striking fact, which is obvious on all figures, is the rise of
the E706 cross sections compared to the theory as $x_{_T}$ decreases
below 0.25. We point out here that this rise is not systematic as
claimed in~\cite{HKKLOT}. In fact, it is not observed at all on the R110
and R807 data and it is rather weak on the R806 data which cover a lower
$x_{_T}$ range than E706. Such a conclusion was also reached by the
authors of~\cite{VV} {\footnote {In \cite{VV} (see also \cite{GORD}) a
better theoretical input than in \cite{HKKLOT}, in particular a complete
NLO treatment of the bremsstrahlung component, was used.}}. 

In conclusion, no set of structure functions is able to accommodate the
recent E706 data despite allowing a reasonable variation in the choice
of scales. In contrast, the NLO calculations are in agreement (within
the rather large experimental and theoretical error bars) with the other
data sets which bracket the E706 data in energy and which, taken
together, cover the large $x_{_T}$ range of E706.

\section{Further phenomenological considerations}
\label{pheno}

In order to bring their data in agreement with theory the E706
collaboration advocates the use of an extra parameter identified as a
measure of the transverse momentum fluctuations of the colliding
partons~\cite{FS}{\footnote{In practice, more than one parameter is needed to
model this effect~\cite{HKKLOT}.}}. In the present phenomenological
implementation of this effect both the normalization and the shape of
the $\pT$ spectrum are affected so that the E706 collaboration is able
to reproduce their data with $k_{_T}$ values in the range 1.2-1.3
GeV/c. For UA6 the choice $k_{_T}=0.7$ GeV/c gives a good agreement with
the data \cite{ABBHMT}. From these values one would expect $k_{_T} > 1.5$ GeV/c to be
appropriate for ISR data which would destroy the rough agreement between
data and theory displayed in the figures. From this we conclude that,
including $k_{_T}$ effects may help some data sets (E706) to agree with
theoretical predictions but it simultaneously destroys the agreement with
other data sets (WA70, ISR) with theory~\cite{FONT}. Globally the phenomenology of
"low" energy photon production is not improved, on the contrary!
More precisely, except for the E706 data there is no need for 
an extra parameter to obtain agreement between data and theory.

Some more comments can be added concerning ``$k_{_T}$" smearing effects.
There is no definite theoretical method to parameterize such effects and
as a result different groups obtain rather different shifts of the
differential cross sections, specially at low values of $\pT$, as it has
been briefly discussed in \cite{ABBHMT}. Besides, the fitted value of
``$k_{_T}$" depends on the estimates used for the perturbative cross
sections. In the low $\pT$ region, the perturbative predictions
become notoriously unstable because no optimum in the factorization and
renormalization scales is achieved. Therefore one observes large
variations in the perturbative estimates under changes of scales. We
comment on this point next.
\begin{figure}[htb]
\begin{center}
\epsfig{file=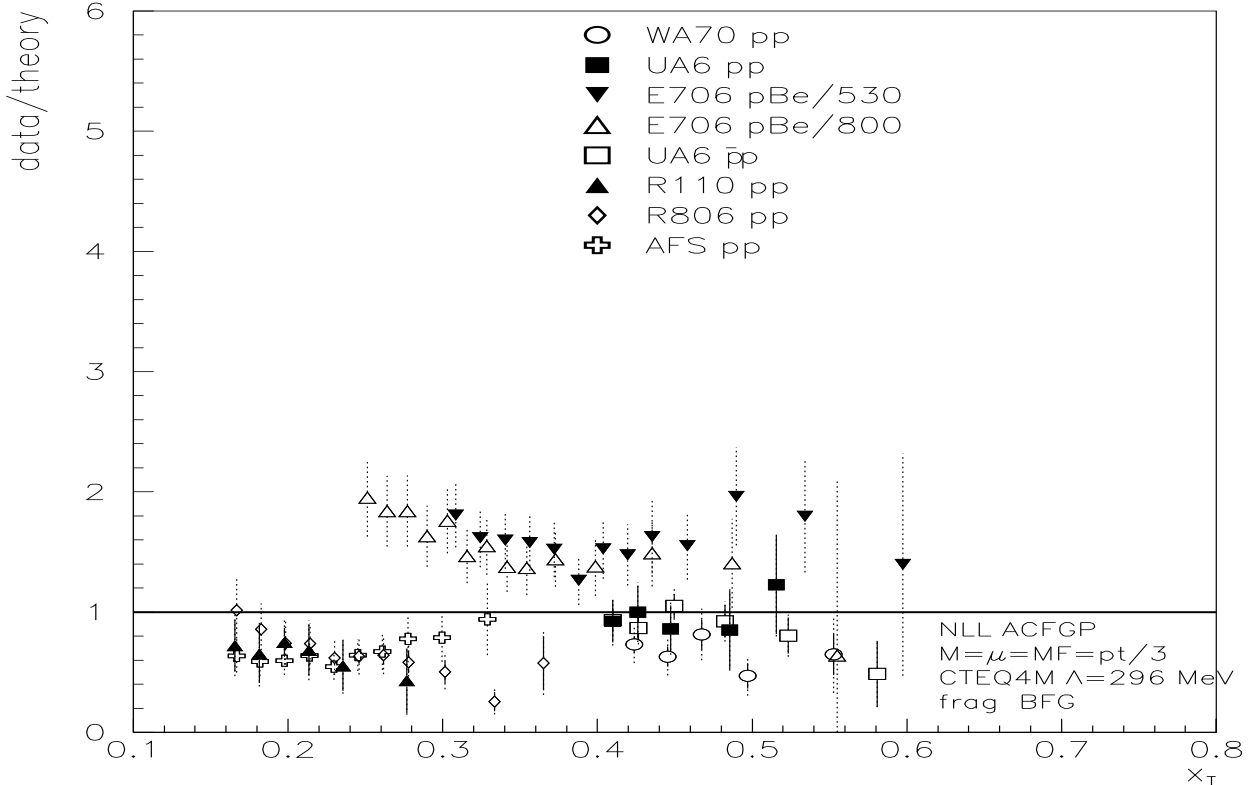,height=12cm,width=18.cm} 
\vskip -1.cm
\caption{\em
Dependence in $x_{_T}$ of the experimental cross sections normalized to
theoretical predictions using the CTEQ4M distributions. All scales in
the calculations have been set to $\pT/3$ and small $\pT$ points have
been dropped as explained in the text.
}
\label{xtcteq4m-1}
\end{center}
\end{figure}
\begin{table}[htb]
\begin{center}
  \begin{tabular}{|c|c|c|c|c|c|c|}
    \hline
     & & & CTEQ4M & CTEQ4M & MRS-98-2& MRS-98-2\\
    &&&&&& \\ 
    Collaboration & Reaction & $\rs$ & {\protect $\chi^2_{_{\pT/2}}$}
 & $\chi^2_{_{\pT/3}}$ & {\protect $\chi^2_{_{\pT/2}}$}
 & $\chi^2_{_{\pT/3}}$  \\
    & & [GeV] & (stat.+syst.) & (stat.+syst.) & (stat.+syst.) & (stat.+syst.)  \\
& & & & & &  \\ \hline \hline
    WA70 \cite{WA70} & $ p\ p$ & 23. & 9.1/7 & 21./5 & 5.7/7 & 37.6/7 \\
\hline
    UA6 \cite{UA6}  & $ p\ p$ &  24.3 & 13.7/9 & 2.34/6 & 19.0/9 & 3.12/8 \\
\hline
    UA6  \cite{UA6} & $ \bar p\ p$ & 24.3 & 5.2/10  & 7.45/7 & 6.44/10 & 5.51/9  \\
\hline
    R806\cite{R806} & $ p\ p$ & 63. & 55.4/11 & 91.6/11 & 91.6/11 & 80.1/11 \\
\hline
    R110\cite{R110} & $ p\ p$ & 63. & 5.7/6 & 14.1/6 & 3.9/6 & 11.2/6 \\  
\hline
AFS/R807\cite{R807} & $ p\ p$ & 63. & 27.8/10 &  71.9/10  & 18.2/10  & 71.6/10  \\
\hline
  \end{tabular}
\end{center}
\caption{\em 
Table of $\chi^2$ values normalized to the number of experimental
points. The statistical errors and systematic errors added in quadrature
are used in the evaluation of $\chi^2$ values. Data points at scales
lower than $Q_0^2$  are not included, which explains why the number of
considered experimental points is smaller for the scale $\pT/3$ than for
the scale $\pT/2$. $Q_0^2=2.56$~GeV$^2$ for CTEQ4M and
$Q_0^2=2.$~GeV$^2$ for MRS-98-2. Only the experimental points with $\pT>
4.2$~GeV/c for WA70/UA6 and with $\pT> 5.$~GeV/c for ISR are kept.
For R110 a $5\%$ uncertainty on the energy scale is not included in the
systematic errors.
} 
\label{chis2-1}
\end{table}
\begin{table}[htb]
\begin{center}
  \begin{tabular}{|c|c|c|c|c|c|c|}
    \hline
     & & & CTEQ4M & CTEQ4M & MRS-98-2& MRS-98-2\\
    &&&&&& \\ 
    Collaboration & Reaction & $\rs$ & {\protect $\chi^2_{_{\pT/2}}$}(stat.)
 & $\chi^2_{_{\pT/3}}$(stat.) & {\protect $\chi^2_{_{\pT/2}}$}(stat.)
 & $\chi^2_{_{\pT/3}}$(stat.)  \\
    & & [GeV] & normaliz. & normaliz. & normaliz. & normaliz.  \\
    & & & & & & \\ \hline \hline
    WA70 \cite{WA70} & $ p\ p$ & 23. & 7.55/7 & 5.88/5 & 9.1/7 & 8.7/7 \\
     & & & $\lambda =$1.46 & $\lambda =$1.27 & $\lambda =$1.21 & $\lambda =$1.61 \\
\hline
    UA6 \cite{UA6}  & $ p\ p$ &  24.3 &5.55/9& 3.4/6  &5.3/9 &3.5/8 \\
     & & & $\lambda =$0.79 & $\lambda =$ 0.98 & $\lambda =$0.74 & $\lambda =$ 0.93\\
\hline
    UA6  \cite{UA6} & $ \bar p\ p$ & 24.3 &6.8/10  & 5.5/7 & 5.9/10&5.5/9  \\
     & & & $\lambda =$0.94 & $\lambda =$1.11 & $\lambda =$0.91 & $\lambda =$1.06 \\
\hline
    E706 \cite{E706} & $ p\ Be$ & 31.6 & 15.7/13 & 12.9/13 & 17.8/13 & 15.1/13 \\
     & & & $\lambda =$0.52 & $\lambda =$0.64 & $\lambda =$0.50 & $\lambda =$0.64 \\
\hline
    E706\cite{E706} & $ p\ Be$ & 38.8 & 64/13 & 51.9/13 & 59.7/13 & 48.7/13  \\
     & & & $\lambda =$0.50 & $\lambda =$0.61 & $\lambda =$ .48 & $\lambda =$0.60 \\
\hline
    R806\cite{R806} & $ p\ p$ & 63. & 329/11 & 278/11 & 348/11 & 281/11 \\
     & & & $\lambda =$0.99 & $\lambda =$1.17& $\lambda =$0.93 & $\lambda =$1.12 \\
\hline
    R110\cite{R110} & $ p\ p$ & 63. & 1.91/6 & 1.62/6 & 1.88/6 & 1.61/6 \\  
     & & & $\lambda =$1.26 & $\lambda =$1.50 & $\lambda =$1.19 & $\lambda =$ 1.43 \\
\hline
AFS/R807\cite{R807} & $ p\ p$ & 63. & 2.6/10 &  3.2/10  & 2.6/10  & 2.7/10  \\
     & & & $\lambda =1.35 $ & $\lambda =1.58$ & $\lambda =1.28$ & $\lambda =1.58$ \\
\hline
  \end{tabular}
\end{center}
\caption{\em 
Table of $\chi^2$ values normalized to the number of experimental
points. The statistical errors are used in the evaluation
of $\chi^2$ values but, for each experiment, an overall normalization
factor is allowed to vary. Only points with $\pT>4.2$ GeV/c for WA70/UA6
and with $\pT>5.$ GeV/c for E706/ISR are kept.
Strictly speaking, $\lambda$ should set to 1 for AFS/R807 since the
quoted errors include the systematics.
}
\label{chis2-2}
\end{table}

We have checked in numerical studies similar to those leading to
fig.~\ref{secmmf} that no stable point, at fixed fragmentation scale,
can be obtained for E706 at 800 GeV and $\pT < 5.$ GeV/c, for example.
The three-dimensional plots show a monotonic variation under changes of
scales as would be obtained in a leading logarithmic calculation. We
would like to interpret this fact as an indication that perturbation
theory is not valid in that range since variation of the arbitrary
scales leads to large variation in the predictions. For quantitative
comparison between theory and experiments such points should not be
included. This phenomenon occurs for the WA70/UA6 data for $\pT$ values
below 4.2 GeV/c. Let us note that at these larger values of transverse
momentum the perturbative predictions are stable against moderate
$k_{_T}$ effects ($<k_{_T}> \ < 0.4$ GeV/c). If we accept this reasonable
limitation the disagreement between data sets and the theoretical
predictions resides essentially in an overall normalization factor and
not in the shape of the $x_{_T}$ distributions as seen in
fig.~\ref{xtmrs982} and fig.~\ref{xtcteq4m-1}.

To gauge the agreement between theory and experiment more quantitatively
and in order to roughly take into account the systematic uncertainties
we proceed in the following way. First, we calculate for each experiment
the $\chi^2$ value per number of experimental points using the
statistical and systematic errors added in quadrature. The results are
shown in table~\ref{chis2-1} except for the E706 data for which
the systematic errors are not yet tabulated. Both sets of structure
functions (CTEQ4M and MRS-98-2) lead essentially to the same
conclusions: no agreement with R806 but satisfactory agreement with the
other experiments. More precisely, the UA6 $\bar p p$ data are in
excellent agreement with both sets of structure functions, independently
of the scale choice which makes them a very good channel to extract the
value of $\Lambda_{_{\overline{MS}}}$~\cite{monique}. The UA6 $p p$ data
coincide very well with both sets of predictions based on the scale
choice $\pT/3$ while clearly the other experiments prefer the larger
scale $\pT/2$ or even a larger scale (alternatively, a lower
$\Lambda_{_{\overline{MS}}}$ value would fit the WA70, R110 and R807
data very well as can be seen from fig.~\ref{rats2}). Turning now to the
E706 data the ratio {\em data/theory} is essentially flat in the $x_{_T}$
region considered as perturbatively reliable (see figs.~\ref{xtmrs982}
and \ref{xtcteq4m-1}) but the data are in excess of the theoretical
prediction by at least $50\%$ both at 530 GeV and 800 GeV. To summarize,
it could be said that all the data sets considered could
be made consistent with  each other (and with the theory) if one would
allow for rather large relative overall normalization shifts.

In table~\ref{chis2-2} we show the result of such a study and give the
$\chi^2$ values based on statistical errors but fitting an overall
normalization factor for each experiment. Clearly, one gets for all
experiments (except R806) very good $\chi^2$ values at the expense of
sometimes unusually large factors $\lambda$. An interesting case is E706:
at 530 GeV the small $\chi^2$ values could have been anticipated from
the figures while at 800 GeV the rather large values displayed in the
table are clearly due to the data points at small $x_{_T}$ with
small statistical errors (but large systematic uncertainties).
Obviously our naive procedure cannot handle such effects correctly. It
would be worthwhile, in this respect, to analyse all data including the
point to point as well as the overall systematic errors properly. We can
just conclude from table~\ref{chis2-2} that the price of obtaining
consistency among all experimental results as well as a good agreement
between theory and experiments is to allow normalization shifts in the
data which are sometimes well in excess of the allowed ranges estimated
from adding the different systematic errors in quadrature. It would
appear that CTEQ4M, with the common scale in the theoretical predictions
set equal to $\pT/3$,  leads to the best phenomenology in the sense that
the normalization factors $\lambda$ are closest to 1, still keeping
small $\chi^2$ values. However, because of the smaller number of
experimental points kept in the analysis it is not possible to argue
decisively in favor of this set rather than MRS-98-2.

It was mentioned above that the shape of the E706 data at 800 GeV  was
not in agreement with the theory at small $x_{_T}$. This raises the
question of how changes of scales affect the shape of the predictions.
This is studied in fig.~\ref{opt} where the theoretical results for
various choices are compared to the reference case with
$\mu=M=M_{_F}=\pT/2$. Two extreme cases are considered, $M_{_F}^2=2$
GeV$^2$ and  $M_{_F}^2=300$ GeV$^2$, the other scales being optimised
in the sense of eq.~(\ref{eq:opt}). The effect of changing the scales
consists essentially in an overall shift in normalization but the
increase in the cross section associated to the small $M_{_F}$ scale is
not enough to bring the theory and the E706 data in agreement. A
relative enhancement is obtained at small $x_{_T}$ with the
(extreme) choice $M_{_F}^2=2$ GeV$^2$ but it is not sufficient to account
for the shape of  the E706/800 GeV data at low $x_{_T}$. A smaller
spread in the theoretical results would, of course, be obtained had we
used $M_{_F}$ scales proportional to $\pT$.  
\begin{figure}[htb]
\begin{center}
\epsfig{file=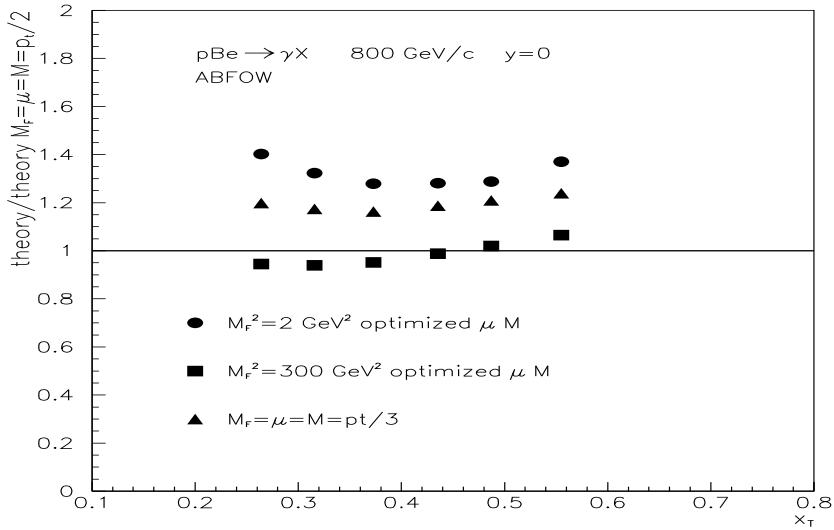,height=8.cm,width=12.cm}
\vskip -.3cm
\caption{\em Study of the sensitivity of the cross section to changes of
scales for E706 at 800 GeV. All results are normalized to the cross
section calculated using $\mu=M=M_{_F}=\pT/2$. 
}
\label{opt}
\end{center}
\end{figure}

A final comment concerning the reliability of the theoretical
predictions is related to the uncertainty in the behaviour of the quark
distributions at large $x$. Recently, it has been pointed out that
nuclear effects in the extraction of the quark distributions from
deep-inelastic data on Deuterium had not been fully taken into account
\cite{YB}. Taking these effects into account  leads to a much larger
$d/u$ ratio at $x>0.5$, increasing with $x$. As a consequence, the
predictions for some prompt photon production rates could be affected.
In fact, $p p$ reactions should not be modified because the photon
production rate is dominated by Compton-like scattering and therefore it
involves the same combination of quark distributions as probed in
deep-inelastic scattering on a proton, namely $(4 u + d)$, which is of
course not sensitive to the effects discussed in \cite{YB}.  Such is not
the case for nuclear targets. For $p N$ scattering, where $N$ is an
isoscalar nucleus, $\sigma^{p N} \sim 5\ (u+d)$ and an increase $\delta
d$ in the $d$ quark distribution manifests itself by an increase $\delta
\sigma^{p N} \sim 15 \delta d /4$ where the constraint $\delta u \sim
-\delta d /4$ has been used. In~\cite{YB} the effect of nuclear binding
is parametrized as an increase of $\delta (d/u) \simeq (.1 \pm .001)
(1+x) x$ of the $d/u$ ratio. This translates into an increase of the
theoretical predictions for $p N$ cross sections
\bea
{\delta\sigma^{p N} \over \sigma^{p N}} \sim {3 \over 4} \ 
	{\delta ({d \over u}) \ (1 - {5 \over 4} {d \over u})}
\ena
which could be as large as $20\%$ when the relevant $<x>$ is large enough.
This is a very rough discussion since, if nuclear effects have to be
disentangled when extracting the $d$ distribution in deep-inelastic
experiments they should be included (back) when making predictions
for prompt photon production on (different) nuclear targets! 
This point certainly deserves further detailed studies.

The predictions for $\bar p p$ could also be modified due to the
important contribution of the annihilation channel $q \bar q \rightarrow
\gamma G$ specially at large $\pT$ where the increase in the $d/u$ ratio
should lead to a slight decrease of the cross section (a quick estimate
leads to $\delta \sigma^{\bar p p}/ \sigma^{\bar p p}|_{ann} \sim
- \delta (d/u) / 2$ for a large enough $\pT$). 

For lack of parton distributions incorporating this effect we do not
pursue a quantitative study of this problem.

\section {Experimental data sets}
\label{datasec}
\setcounter{footnote}{0}
In view of possible incompatibilities between various
data sets, let us review some specific features  of the 
experiments.
Details on systematics uncertainties are not always available.
For the latest E706 data, details on energy scale
uncertainty are available. For other issues, we comment on the early
data analysis \cite{E70693} which may not apply for the large statistical
sample. It would be interesting to have such information  on the large
statistical sample for a detailed comparison between data sets.

\subsection{Hydrogen or heavier nuclei}
In fixed target experiments, the data mainly have been taken at CERN 
with hydrogen
(with the exception of NA3 data on isoscalar Carbon target) while
the  published direct photon cross sections
with heavier nuclei come mainly from the E706 collaboration \cite{E70693,E706}.
A detailed study of the nuclear corrections is not available.

\subsection{Isolation criteria}
All the comparisons with theory have assumed fully inclusive
photon production. In fact, the prompt photon samples are
always obtained with some isolation criteria, at least to remove
electrons and charged hadrons in the sample. These cuts depend on
the granularity 
of the detectors which is an important ingredient of the  shower 
reconstruction algorithms.
The data presented are usually corrected for this inefficiency. 
\par
These cuts were sizable for the pioneering ISR
experiments.
In the
R110 sample, where the cuts were no charged tracks within 25 cm
of the shower and no additional electromagnetic shower within
20 cm of the candidate photon,
no correction was applied and the data are presented 
as biased against bremsstrahlung \cite{R110}. A proper theoretical
treatment of such cuts is not available.
The R806 \cite{R806} and AFS \cite{R807} data, on the
other hand, are corrected for such effects. The correction is 1.15 to 1.25 
for the later but no details are given on the $\pT$ dependence 
of the correction. Part of the observed discrepancies between the
ISR experiments at low $\pT$ could be attributed to this effect.
\par
For fixed target data, two kinds of cuts are applied,
one on a charged track extrapolating to the shower of
the photon candidate to remove electrons and charged
hadron showers, and one on the longitudinal energy deposition
to remove hadronic showers in the sample.
For WA70 and UA6, the cut is no charged track 
extrapolating within 5 cm, respectively 1.5 cm, of the shower center.
For E706, no details are given for the present data, but earlier
data were obtained without charged track extrapolating
to 1~cm of the shower center \cite{E70693}.
For UA6 and E706 (530), due to the difference in energy, these cuts
cover about the same phase space while for WA70 the cut is more drastic.
Cuts on the longitudinal energy deposition are implemented
differently by WA70 (at the shower tail) and E706 (at the shower front)  
and are corrected for.
These two kinds of cuts have more effects on bremsstrahlung
than direct photon production, {\em i.e.} more effects at low $\pT$.
Small differences on cross sections may result from the different cuts
and the different ways their efficiencies are estimated.
They are surely bounded by the expected size of the bremsstrahlung contribution
(which should not exceed 20\%).

\subsection{Backgrounds} 
The experimental candidate photon samples are always
contaminated by sizable backgrounds. With high granularity
calorimeters such as those used by WA70, UA6 and E706, the
backgrounds are mainly from $\pi^0$. 
The amount of background depends on the calorimeter, 
for example granularity, and on the strategies
adopted to reconstruct the showers. It  obviously depends on 
$\pT$, on rapidity, and is important at low $\pT$ in any case.
The background corrections are performed 
with simulation of the known backgrounds in the detector, the simulation
being checked on the reconstructed $\pi^0$'s.

Typical background fractions in the candidate photon sample
are of order 50-60\% at $\pT$= 4.25 GeV/c to 20-30\% at $\pT$=5.5 GeV/c 
(WA70, central rapidity);
70 to 80\% at $\pT$ of 4-5 GeV/c (E706 early data). 
For UA6 $\bar{p}p$ data, the corresponding figure is about 40\% 
and depends on $\pT$ while for $pp$ it depends weakly on $\pT$ and 
amounts to about 50\% \cite{UA6thesis}.
Typical uncertainties on the cross section quoted are 16\% at $\pT$=4 GeV
down to 8\% at $\pT$ of 8 GeV/c for E706 early data. For WA70, the
uncertainties were in the 10-25\% range with an additional 15\% 
upper limit from simulation. UA6 quotes specifically the various 
independent uncertainties on simulation in the 1 to 6\% range. 

Small errors on  background evaluation induces large errors on the
extracted signals. It is therefore very important to understand the background
precisely. This point deserves futher study as it may be helpful
to understand discrepancies between experiments.  

\subsection{Comparison of data sets}
Without theoretical prejudice, one can only compare data in a similar
energy range. For the CERN fixed target data   from the WA70 (beam of
280 GeV/c \cite{WA70}) and the NA24 (beam of 300 GeV/c) experiments,
``the agreement  (at y=0) appears to be well within systematics and even
statistical uncertainty" \cite{Ferbel,Martin}. On the other hand, at ISR
with $\sqrt{s}=62.3$ GeV, ``normalizations were about 1.5 to 2 in relative
doubt"~\cite{Ferbel} {\footnote{As for the published ISR data used in this
present work, the discrepancy is in the slope in $\pT$ rather than in
the overall normalization.}}.

Although most of the ISR prompt photon data have been taken at $\rs$=63 GeV
\cite{R806,R110,R807}, 
data at lower $\sqrt{s}$ 
{\footnote{Data on $\gamma/$all neutral clusters
have been reported at $\sqrt{s}=44.8$ GeV and 62.4 GeV by the R110
collaboration \cite{R110gp}.
At fixed $\pT$, the $\gamma/$all neutral clusters decreases with
$\sqrt{s}$~\cite{R110gp}  while $\gamma/\pi^0$ shows little $\sqrt{s}$ 
dependence \cite{R806}.}}
are reported  as 
$\gamma/\pi^0$ by the R806 collaboration (at $\sqrt{s}=31$ GeV, 45 GeV
and 63 GeV \cite{R806,diak80}).
One can tentatively combine the data on $\gamma/\pi^0$ \cite{diak80}
at $\sqrt{s}=31$ GeV with
$\pi^0$ production  cross sections (using the super-retracted data 
from table 5 of Ref. \cite{kour80} at $\sqrt{s}=30.6$ GeV)
{\footnote {These $\pi^0$ cross sections are in fair
agreement with measurements of the R110 collaborations
at $\sqrt{s}=31$ GeV \cite{R110PI}.}}: 
the resulting $\gamma$ cross sections 
turn out to be, within large experimental errors, compatible
with the E706 results in the $\pT$ range $4.0-6.0$ GeV/c.
It is worth mentioning that this is just a rough cross-check
as the R806 collaboration did not publish a direct photon cross section at
$\rs=31$ GeV
{\footnote{As a check of this rough procedure,
combining the same way data on  $\gamma/\pi^0$ at $\sqrt{s}=63$ GeV
\cite{diak80}
with the super-retracted data at $\sqrt{s}=62.8$ GeV of Ref. \cite{kour80}
one gets photon cross sections statistically compatible 
with Ref. \cite{R806}.}}.

We have not included these data on the {\em data/theory} plots as they
do not help to solve the observed discrepancies on these plots  between
the rough agreement with perturbative theory of the bulk of ISR data at
low $x_{_T}$ and the strong disagreement of the E706 data at higher
$x_{_T}$: within the deduced experimental errors, these ratios for the
data at $\rs=31$ GeV  are compatible with the E706 data at $\rs=31.6$
GeV in the $x_{_T}$ range $0.26-0.39$  but they are also compatible with
1. Unfortunately we therefore cannot conclude on ISR/E706 comparisons.

\subsection{Summary on experimental data sets}
Other effects, such as luminosity, trigger efficiencies, absolute
energy scale, may affect the $\gamma$ cross section. However
they also affect the $\pi^0$ cross section. It is more
relevant to discuss them in this context \cite{nous}.

\section {Conclusions}

It appears that the present inclusive prompt photon data at fixed target
and ISR energies are incompatible: normalizing the theory on one set, the
extrapolation of the theory to a slightly different energy completely
misses the corresponding data. It does not appear very instructive to
hide this problem by introducing an extra parameter fitted to the data
at each energy: the agreement with some data sets is improved while the
agreement with other sets is destroyed.  Besides, the actual value of
the extra parameter depends on the value chosen for the scales
used in the perturbative predictions. Since, anyhow, the perturbative
predictions become unstable at low $\pT$ where the ``intrinsic"
momentum contributes most, the extracted value of this parameter suffers
from a large error. It seems more reasonable to us to only use data at
reasonably large $\pT$ values. In this region, the various sets appear
to show less important discrepancies. In any case, an effort should be
done by experimentalists to extend the discussion of sec.~\ref{datasec}
to better understand the various sources of systematic uncertainties.
Perhaps RHIC, with a center-of-mass energy  as low as 50 GeV,
could help resolve the important discrepancy observed between some
data sets. 

As a final remark, let us notice that all direct photon experiments also
measure inclusive $\pi^0$ cross sections. It would be interesting to see
whether the normalization discrepancies observed in
direct photon experiments
%
%
persist for the $\pi^0$ cross sections. Here the theoretical predictions
are less reliable: 1) the cross sections depend on the fragmentation
functions of quarks and gluons into $\pi^0$; 2) the effective value of
the fragmentation variable $z$ is in the range .8-.9 and large $\ln
(1-z)$ terms should be resummed; 3) the scale dependence of the cross
section is large. For all these reasons, we do not expect a very good
agreement between data and theory in the $\pT$-range of fixed target
experiments~\cite{C..W}. We can however use the theoretical cross
sections as reference curves and investigate how data are situated with
respect to the predictions. Preliminary studies~\cite{nous} show that
fixed target data overshoot the theoretical curves by a few tens of
percents, when CTEQ4M or MRST and the scales $\pT/3$ are used. Therefore
all fixed target $\pi^0$ data display a similar behaviour compared to
the theoretical cross sections and appear less scattered than the prompt
photon data. This would indicate that the main problem rests with the
extraction of the $\gamma/\pi^0$ ratio.

\section {Acknowledgements}

We thank L. Camilleri, J. Huston, G. Ginther and M. Zielinski for discussions.



\begin{thebibliography}{99}
%
\bibitem{EXP}	E629 collaboration, M.~McLaughlin {\em et. al.}, \PRL
		{51} (1983) 971;\\
		NA3 collaboration, J.~Badier {\em et. al.}, \ZPH {C31}
		(1986) 341;\\
		NA24 collaboration, C.~De Marzo {\em et. al.}, \PR {D36}
		(1987) 8;\\
		R108 collaboration, A.L.S.~Angelis {\em et. al.}, 
		\NP {B263} (1986) 228; \\
		E704 collaboration, D.~Adams {\em et al.}, \PL {B345}
		(1995) 569.
\bibitem{R806}	R806 collaboration,  E.~Annassontzis {\em et. al.}, \ZPH
		{C13} (1982) 277.
\bibitem{WA70}	WA70 collaboration, M.~Bonesini {\em et al.}, \ZPH {C37}
		(1988) 535.
\bibitem{WA70pi}WA70 collaboration, M.~Bonesini {\em et al.}, \ZPH {C38}
		(1988) 371.
\bibitem{R110}	R110 collaboration, A.L.S.~Angelis {\em et. al.}, 
		\NP {B327} (1989) 541.
\bibitem{R807}	AFS/R807 collaboration,  T.~\AA kesson {\em et. al.}, 
		Sov. J. Nuc. Phys. {\bf 51} (1990) 836.
\bibitem{E70693} E706 collaboration, G.~Alverson {\em et al.},
		\PR {D48}(1993) 5.
\bibitem{E706}	E706 collaboration, L.~Apanasevich {\em et al.}, 
		\PRL {81} (1998) 2642.
\bibitem{UA6}	UA6 collaboration, G.~Ballocchi {\em et al.}, 
		\PL {B436} (1998) 222.
\bibitem{VW}	For a compilation of data, see W. Vogelsang and 
		M.R. Whalley, J. Phys. {\bf G23}, Suppl. 7A (1997) A1.
\bibitem{FRIT}	H.~Fritzsch and P.~Minkowski, \PL {71B} (1977) 392.
\bibitem{ABDFS}	P.~Aurenche, R.~Baier, M.~Fontannaz and D.Schiff, \NP
		{B297} (1988) 661.
\bibitem{ACFGP}	P.~Aurenche, P.~Chiappetta, M.~Fontannaz, J.Ph.~Guillet,
		and E.~Pilon, \NP {B399} (1993) 34.
\bibitem{GV}	L.E.~Gordon and W.~Vogelsang, \PR {D50} (1994) 1901.
\bibitem{GRUN}	G.~Grunberg, \PR {D29} (1984) 2315.
\bibitem{STEV}	P.M.~Stevenson, \PR {D23} (1981) 2916;\\
		P.M.~Stevenson and H.D.~Politzer, \NP {277} (1986) 758.
\bibitem{BLM}	S.J.~Brodsky, G.P.~Lepage and P.B.~Mackenzie,
		\PR {D28} (1983) 228. 
\bibitem{OPT}	P.~Aurenche, R.~Baier, M.~Fontannaz and D.Schiff, \NP
		{B286} (1987) 509.
\bibitem{ABFOW}	P.~Aurenche, R.~Baier, M.~Fontannaz, J.F.~Owens and
		M.~Werlen, \PR {D39} (1989) 3275.
\bibitem{VV}	W. Vogelsang and A. Vogt, \NP {B453} (1995) 334. 
\bibitem{GORD}	L.E.~Gordon, \NP {B501} (1997) 175.
\bibitem{HKKLOT}J. Huston, E. Kovacs, S. Kuhlmann, H.L. Lai, J.F. Owens
		and W.K. Tung, \PR {D51} (1995) 6139.  
\bibitem{FERB}	T.~Ferbel, {\em in} proceedings of the 1996 Rencontres de Moriond,
		hadronic session, J. Tr\^an Thanh V\^an ed., Editions Fronti\`eres.
\bibitem{ABBHMT}L.~Apanasevich {\em et al.}, preprint CTEQ-805,
		hep-ph/9808467. 
\bibitem{ZIEL}	M.~Zieli\'nski, hep-ph/9811278.
\bibitem{ACGG}	F.~Aversa, P.~Chiappetta, M.~Greco and J.Ph.~Guillet,
		\NP {B327} (1989) 104.
\bibitem{CTEQ}	CTEQ collaboration, H.L.~Lai {\em et al.}, \PR {D55}
		(1997) 1280.
\bibitem{MRS}	A.D.~Martin, R.G.~Roberts and W.J.~Stirling and R.S.~Thorne,
                Eur. Phys. J. \EPJ {C4} (1998) 463.
\bibitem{GRVdis}M. Gl\"uck, E. Reya and A. Vogt, \EPJ {C5} (1998) 461.
\bibitem{GRV}	M. Gl\"uck, E. Reya and A. Vogt, \PR {D48} (1993) 116;
		Erratum: {\em ibid.} {\bf D51},1427.
\bibitem{BFG}	L.~Bourhis, M.~Fontannaz and J.Ph.~Guillet, \EPJ {C2}
        	(1998) 529.
\bibitem{CMN} 	E.~Laenen, G..Oderda and G..Sterman, hep-ph/9806467;\\
		S.~Catani, M.L.~Mangano and P.~Nason, 
   		J. High Energy Phys. (1998) 9807:024.                             
\bibitem{FS}	M.~Fontannaz and D.~Schiff, \NP {B132} (1978) 457.
\bibitem{FONT}	M.~Fontannaz {\em in} proceedings of the 1997 Rencontres
		de Moriond, hadronic session, J. Tr\^an Thanh V\^an 
		ed., Editions Fronti\`eres.
\bibitem{monique}UA6 collaboration, G.~Ballocchi {\em et al.},
    		Phys. Lett. {\bf B317} (1993) 250;\\
    		M.~Werlen {\em in} Proc. Int. Europhysics Conf. on High Energy
		Physics, EPS 93, Marseille, ed. by J.~Carr and 
		M.~Perrottet, (Editions Fronti\`eres) 323; \\
                UA6 Collaboration, G.~Ballocchi {\em et al.}, 
		in preparation.
\bibitem{YB}	U.K.~Yang, A.~Bodek and Q.~Fan, {\em in} proceedings of 
		the 1998 Rencontres de Moriond, hadronic session, 
		J. Tr\^an Thanh V\^an ed., Editions Fronti\`eres.
\bibitem{UA6thesis} see Ph.~D. theses D. Hubbard (Michigan 1993),
                P.~Oberson (Lausanne 1994), C.~Comtat (Lausanne 1996), 
		N.M.~Chung (Lausanne 1998).
\bibitem{Ferbel} T.~Ferbel {\em in} Proc. Physics in Collisions, Capri 
		1988, 255, P.~Strolin ed., Editions Fronti\`eres..
\bibitem{Martin} M.~Martin {\em in} Proc. QCD Workshop, St. Croix 1988,
                 B.~Cox ed., Plenum Pub. Corp.
\bibitem{R110gp} R110 collaboration, A.L.S.~Angelis {\em et al.},		
		\PL {98B} (1981) 115.	  		 
\bibitem{diak80} R806 collaboration, M.~Diakonou {\em et al.},
		\PL {91B} (1980) 296.
\bibitem{kour80}R806 collaboration, C.~Kourkoumelis {\em et al.},
		\ZPH {C5} (1980) 5.	 
\bibitem{R110PI} R110 collaboration, A.L.S.~Angelis {\em et al.},	
		\PL {185B} (1987) 213.
\bibitem{nous} 	P.~Aurenche, M.~Fontannaz, J.Ph.~Guillet, B. Kniehl,
		E.~Pilon and M.~Werlen, preprint in preparation.
\bibitem{C..W} 	P.~Chiappetta, M.~Greco, J.Ph.~Guillet, S.~Rolli and
		M.~Werlen, \NP {B412} (1994) 3.


\end{thebibliography}
\end{document}